\documentclass[onecolumn,superscriptaddress,preprintnumbers,nofootinbib,notitlepage]{revtex4}

\usepackage{bm}
\usepackage{feynmp}
\usepackage{simplewick}
\usepackage{graphics,graphicx}
\usepackage{amsmath,amssymb}
\usepackage{color}
\usepackage{array,mathtools}
\usepackage{booktabs,multirow}
\usepackage{bbold}
\usepackage[normalem]{ulem}
\usepackage[colorinlistoftodos,prependcaption,textsize=tiny]{todonotes}		
\usepackage{braket}
\usepackage{pgfplots}

\newcommand{\bea}{\begin{eqnarray}}
\newcommand{\eea}{\end{eqnarray}}

\newcommand{\beq}{\begin{equation}}

\newcommand{\eeq}{\end{equation}}

\newcommand{\MeV}{{\rm MeV}}
\definecolor{nred}{RGB}  {240, 50,  0}
\definecolor{nblue}{RGB} {  0, 50,200}
\definecolor{ngreen}{RGB}{ 50,200,  0}

\makeatother

\begin{document}
\preprint{\tt LPT Orsay 19-33}

\vspace*{22mm}

\title{Decay constant of $B_s$ and $B^*_s$ mesons from ${\rm N_f}=2$ lattice QCD}

\author{Rahul~Balasubramamian}
\affiliation{Laboratoire de Physique Th\'eorique\footnote[1]{Unit\'e Mixte de Recherche 8627 du Centre National de la Recherche Scientifique}, CNRS, Univ. Paris-Sud et Universit\'e Paris-Saclay,  B\^atiment 210,  91405 Orsay Cedex, France}\affiliation{Nikhef, Science Park 105, NL-1098 XG Amsterdam, The Netherlands},\author{Beno\^it~Blossier}
\affiliation{Laboratoire de Physique Th\'eorique\footnote[1]{Unit\'e Mixte de Recherche 8627 du Centre National de la Recherche Scientifique}, CNRS, Univ. Paris-Sud et Universit\'e Paris-Saclay,  B\^atiment 210,  91405 Orsay Cedex, France}

\begin{abstract}
We report on a two-flavor lattice QCD estimate of the $B_s$ and $B^*_s$ leptonic decays parameterized by the decay constants $f_{B_s}$
and $f_{B^*_s}$. In addition to their relevance for phenomenology, their extraction has allowed us to investigate whether the ``step scaling 
in mass" strategy is suitable with Wlilson-Clover fermions to smoothly extrapolate quantities of the heavy-strange sector up to the bottom scale.
From the central value of $f_{D_s}$ quoted by FLAG at $N_f=2$ and our ratio $\frac{f_{B_s}}{f_{D_s}}$,
we obtain $f_{B_s}=215(10)(2)(^{+2}_{-5})$ MeV and $f_{B^*_s}/f_{B_s}=1.02(2)(^{+2}_{-0})$.
\end{abstract}

\maketitle

\section{\label{Introduction}Introduction}

In the very active research of new effects in high-energy particle physics, flavour
physics does play a key role at the so-called intensity frontier.  Indeed, rare events are
sensitive probes of New Physics (NP) scenarios with the exchange of extra particles in quantum loops
with respect to what is known from the Standard Model (SM),  However, theoretical uncertainties on
hadronic quantities, for instance hadron decay constants, that encode the dynamics of QCD at large distance, severely weaken 
the constraints that are derived through the analysis of experimental data. Those hadronic constants
cannot be reliably estimated in perturbation theory. $b$-quark physics is a particularly interesting place to search for NP effects and it has recently regained even stronger
attention after experimental signs of several anomalies in $B$ and $B_c$ decays. More precisely several ratios $R_D=\frac{\Gamma(B \to D \tau \nu_\tau)}
{\Gamma(B \to D \ell \nu_\ell)_{\ell = e,\mu}}$, $R_{D^*}=\frac{\Gamma(B \to D^* \tau \nu_\tau)}
{\Gamma(B \to D^* \ell \nu_\ell)_{\ell = e,\mu}}$, $R_{J/\psi}=\frac{\Gamma(B_c \to J/\psi \tau \nu_\tau)}
{\Gamma(B_c \to J/\psi \ell \nu_\ell)_{\ell = e,\mu}}$, $R_K=\frac{\Gamma(B \to K \mu^+ \mu^-)}
{\Gamma(B \to K e^+ e^-)}$ and $R_{K^*}=\frac{\Gamma(B \to K^* \mu^+ \mu^-)}
{\Gamma(B \to K^* e^+ e^-)}$ show some discrepancy with SM expectations \cite{LeesXJ} -- \cite{BifaniZMI}. The three former might bring stringent constraints on $\bar{b}c$ currents,
for instance mediated by the exchange of 
leptoquarks \cite{DiLuzioVAT}. The further ratios $R_{D^{(*)}_s}=\frac{\Gamma(B \to D^{(*)}_s \tau \nu_\tau)}{\Gamma(B \to D^{(*)}_s \ell \nu_\ell)_{\ell = e,\mu}}$, under investigation at LHCb, will provide even more informations once, on the theory side, the hadronic matrix elements associated to $B_s \to D^{(*)}_s$ are under comparable control by means of lattice QCD. Simulating the $B_s$ meson on the lattice is delicate as far as cut-off effects are concerned. Several strategies have been followed in the literature, including simulations of relativistic $b$-quarks using an action tuned so as to minimize
discretization errors \cite{McNeileNG} -- \cite{ChristUS}, the use of Non Relativistic QCD \cite{NaKP}, \cite{DowdallTGA}, performing computations in Heavy Quark Effective Theory  (HQET) \cite{BlossierMK} and the extrapolation of
simulation results obtained in the region between the charm quark mass $m_c$
and a mass $\sim 3 m_c$ to the physical $b$-quark mass
\cite{DimopoulosGX}, \cite{CarrascoZTA}. As we plan to employ the latter approach to study $B_s$ decays with $O(a)$ improved Wilson-Clover fermions, an intermediate step is to extract $f_{B_s}$ and $f_{B^*_s}$, in order to validate the method. The lattice QCD community has made a significant effort to compute $f_{B_s}$ with ${\rm N_f=2}$ \cite{CarrascoZTA}, \cite{BernardoniFVA}, ${\rm N_f}=2+1$ \cite{McNeileNG}, \cite{BazavovAA}, \cite{NaKP}, \cite{AokiNGA}, \cite{ChristUEA} and ${\rm N_f}=2+1+1$ \cite{DowdallTGA}, \cite{BussoneIUA} -- \cite{HughesSPC}. Recently, the SU(3) symmetry breaking $f_{B_s}/f_B$ has been extracted at the physical point \cite{BoyleKNM}. Concerning the spin-symmetry breaking ratio $f_{B^*_s}/f_{B_s}$ only 2 lattice results are available, both at ${\rm N_f}=2+1+1$ \cite{ColquhounOHA}, \cite{LubiczASP}. Ratios $f_{B^*}/f_B$ and $f_{B^*_s}/f_{B_s}$ have been investigated with other methods than lattice simulations, i.e. constituent quark models \cite{MelikhovYU}, \cite{EbertHJ} and QCD sum rules \cite{GelhausenWIA} – \cite{WangMXA}.

The paper is organized as follows: in section \ref{sec2} we recall what is the ``step scaling in mass" strategy, in section \ref{sec3}
we present the simulations details and our raw data, and in section \ref{sec4} we describe our analysis and comment the results. Finally we conclude in section \ref{sec5}.

\section{Step scaling in mass with Wilson-Clover fermions\label{sec2}}
 
The idea is to extract $f_{B_s}\equiv \frac{f_{B_s}}{f_{D_s}} \times f_{D_s}$ and $f_{B^*_s}\equiv \frac{f_{B^*_s}}{f_{B_s}} \times f_{B_s}$ by separate measurements, of the quantity 
$f_{D_s}$ on one side, the ratio $\frac{f_{B_s}}{f_{D_s}}$ on another side and the ratio $\frac{f_{B^*_s}}{f_{B_s}}$ on a third side. In the following we focus on the two latter in the framework of $N_f=2$ Wilson-Clover fermions. With a given pion mass, through $\kappa_{\rm sea}$, the valence strange quark mass $\kappa_s$ and lattice spacing $a$, we consider the ratio
\begin{equation}
\label{eqratfp}
\frac{[f_P \sqrt{M_P}](a,\kappa_{\rm sea},\kappa_{h_{i+1}},\kappa_s)C^{\rm stat}_A(\mu_f, M_P(a,\kappa_{\rm sea},\kappa_{h_{i}},\kappa_s))}{[f_P \sqrt{M_P}](a,\kappa_{\rm sea},\kappa_{h_i},\kappa_s)C^{\rm stat}_A(\mu_f, M_P(a,\kappa_{\rm sea},\kappa_{h_{i+1}},\kappa_s))}\frac{1}{\sqrt{\lambda}} \equiv r_P(a,\kappa_{\rm sea},\kappa_{h_i},\kappa_{h_{i+1}},\kappa_s).
\end{equation}
where $\lambda=\left(\frac{m_{B_s}}{m_{D_s}}\right)^{\frac{1}{K}}$, $i=0,\cdots,K-1$, $\kappa_{h_i}$ is a valence heavy quark mass, and $M_P(a,\kappa_{\rm sea},\kappa_{h_0},\kappa_s) \equiv m_{D_s}$, up to mistuning effects. 
For later usage it is convenient to redefine $r_P$ as
\begin{eqnarray}
\nonumber
r_P(a,\kappa_{\rm sea},\kappa_{h_i},\kappa_{h_{i+1}},\kappa_s) &\equiv&
\frac{C^{\rm stat}_A(\mu_f, M_P(a,\kappa_{\rm sea},\kappa_{h_i},\kappa_s))}
{C^{\rm stat}_A(\mu_f, M_P(a,\kappa_{\rm sea},\kappa_{h_{i+1}},\kappa_s))} r'_P(a,\kappa_{\rm sea},\kappa_{h_i},\kappa_{h_{i+1}},\kappa_s),\\
r'_P(a,\kappa_{\rm sea},\kappa_{h_i},\kappa_{h_{i+1}},\kappa_s)&=&
\frac{f_P(a,\kappa_{\rm sea},\kappa_{h_{i+1}},\kappa_s)}{f_P(a,\kappa_{\rm sea},\kappa_{h_i},\kappa_s)}.
\end{eqnarray}
$C^{\rm stat}_A(\mu_1,\mu_2)$ and $C^{\rm stat}_V(\mu_1,\mu_2)$, that will appear later in the paper, are the matching coefficients between the QCD currents $J^{\rm QCD}_{A(V)}\equiv \bar{Q}\gamma_0\gamma^5 q(\bar{Q} \gamma_1 q)$ and their HQET counterpart $J^{\rm HQET}_{A(V)}\equiv\bar{h}\gamma_0 \gamma^5 q(\bar{h} \gamma_1 q)$ defined at the renormalization scale $\mu_1$, $J^{\rm HQET}_{A(V)}(\mu_1)=C^{\rm stat\,-1}_{A(V)}(\mu_1,\mu_Q)J^{\rm QCD}_{A(V)}$, where $\mu_Q$ is a scale related to the heavy quark mass $m_Q$, for instance the heavy-light pseudoscalar meson mass $M_P(Q, q)$. $C^{\rm stat}_A$ and $C^{\rm stat}_V$ are known up to 3-loop of perturbation theory \cite{BroadhurstSE} -- \cite{BekavacZC}\footnote{In this work we have taken the N$^2$LO formulae for the matching between QCD and HQET at the scale $\mu_Q$.}. $r_P$ is independent of $\mu_f$ because it involves the renormalization group equation of $C^{\rm stat}_A$ integrated from $M_P(a,\kappa_{\rm sea},\kappa_{h_i},\kappa_s))$ to $M_P(a,\kappa_{\rm sea},\kappa_{h_{i+1}},\kappa_s)$.
Thanks to scaling laws in HQET, it is expected that $r_P$ has a simple expansion in the inverse heavy quark mass defined in a specific renormalization scheme, for instance the pole mass 
\cite{BlossierHG}. But,  in the case of Wilson-Clover fermions and contrary to the case of twisted-mass fermions, there is so far no straightforward relation between the pole quark mass and the bare quark mass, through the renormalization group invariant (RGI) quark mass, \emph{if the quark is significantly heavier than the charm}. Indeed,  using the series of RGI masses $m^{\rm RGI}_{h_i}$ such that $m^{\rm RGI}_{h_i}/m^{\rm RGI}_{h_{i+1}}=1/\lambda'$, $m^{\rm RGI}_b=\lambda'^K m^{\rm RGI}_c$ with $m^{\rm RGI}_b$ already determined \cite{BernardoniXBA} and $m^{\rm RGI}_c$  known after the tuning of $\kappa_c$, we define the \emph{improved} RGI mass $m^{\rm RGI}_{h_i} \propto (1+b_m a m_{h_i})m_{h_i}$, where $m_{h_i}$ is the heavy vector Ward Identity quark mass $m_{h(i)}=\frac{1}{2\kappa_{h_i}}-\frac{1}{2\kappa_{\rm critical}}$. The problem is that we can have $1+b_m a m_{h_i}<0$ because the improvement coefficient $b_m$ is negative \cite{FritzschAW}. Negative RGI masses are of course not physical. The issue would be solved by adding the ${\cal O}(a^2)$ term in the definition of the RGI mass, which is unfortunately unknown. That is why we have decided to consider the inverse of pseudoscalar heavy-strange pseudoscalar meson masses $M_P(\kappa_{h_i},\kappa_s) \equiv M_{H_s}(i)$ as the parameter expansion of $r_P$ and we define the steps as sequential ratios $\frac{M_P(\kappa_{h_{i+1}},\kappa_s)}{M_P(\kappa_{h_i},\kappa_s)}$ that should be constant with the regulator $a$
and the sea quark mass $\kappa_{\rm sea}$. They are the analogous of the ratios of RGI quark masses $\frac{m^{\rm RGI}_{h_{i+1}}}{m^{\rm RGI}_{h_i}}$.

\section{Lattice calculation details\label{sec3}}

\renewcommand{\arraystretch}{1.2}
\setlength{\tabcolsep}{8pt}
\begin{table}[t]
\begin{center}
\begin{tabular}{lcc@{\hskip 02em}c@{\hskip 02em}c@{\hskip 01em}c@{\hskip 01em}c@{\hskip 01em}c@{\hskip 01em}c@{\hskip 01em}c}
\hline
	\toprule
	id	&	$\quad\beta\quad$	&	$(L/a)^3\times (T/a)$ 		&	$\kappa_{\rm sea}$		&	$a~(\rm fm)$	&	$m_{\pi}~(\MeV)$	& $Lm_{\pi}$ 	& $\#$ cfgs&$\kappa_s$&$\kappa_c$\\
\hline 
	\midrule
	A5	&	5.2		&	$32^3\times64$	& 	$0.13594$	& 	0.0751	  	& 	$333$	&4.1	& $198$&$0.135267$&$0.12531$\\  
	B6	&			& 	$48^3\times96$	&	$0.13597$	& 			& 	$282$	&5.2	& $126$&$0.135257$&$0.12529$\\    
\hline 
	\midrule
	E5	&	5.3		&	$32^3\times64$	& 	$0.13625$	& 	0.0653	  	& 	$439$	&4.7	& $200$&$0.135777$&$0.12724$\\  
	F6	&			& 	$48^3\times96$	&	$0.13635$	& 			& 	$313$	&5	& $120$&$0.135741$&$0.12713$\\    
	F7	&			& 	$48^3\times96$	&	$0.13638$	& 			& 	$268$	&4.3	& $200$&$0.135730$&$0.12713$\\    
	G8	&			& 	$64^3\times128$	&	$0.13642$	& 			& 	$194$	&4.1	& $176$&$0.135705$&$0.12710$\\    
\hline
	\midrule
	N6	&	$5.5$	&	$48^3\times96$	&	$0.13667$	& 	$0.0483$  	& 	$341$	&4	& $192$&$0.136250$&$0.13026$\\	
	O7	&		&	$64^3\times128$	&	$0.13671$	& 	 	& 	$269$	&4.2	& $160$&$0.136243$&$0.13022$ \\ 
	\bottomrule
\hline
\end{tabular} 
\end{center}
\caption{Parameters of the simulations: bare coupling $\beta = 6/g_0^2$, lattice resolution, hopping parameter $\kappa$, lattice spacing $a$ in physical units, pion mass, number of gauge configurations, bare strange and charm quark masses.}
\label{tabsim}
\end{table}

\begin{figure}[t]
\begin{center}
\includegraphics*[width=9cm, height=5cm]{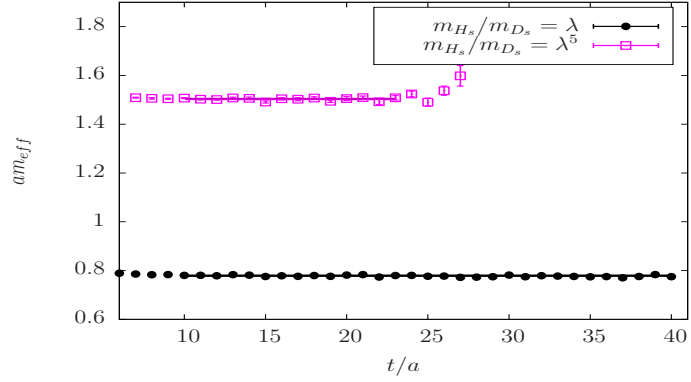}
\end{center}
\caption{\label{fig:platmas} Effective mass of two pseudoscalar heavy-strange mesons extracted with the projected 2-pt correlation function $\tilde{C}_{A_0P}$ along the the generalised eigenvector $v^1_P(t_{\rm fix},t_0)$. The CLS ensemble is F7.}
\end{figure}

We have performed our analysis from the CLS ensembles made of $N_f=2$ nonperturbatively $\mathrm{O}(a)$-improved Wilson-Clover fermions 
\cite{SheikholeslamiIJ, LuscherUG} and the plaquette gauge action \cite{WilsonSK}. 
In Table~\ref{tabsim} we collect the main informations about the simulations. Three lattice spacings $a_{\beta=5.5}=0.04831(38)$ fm, $a_{\beta=5.3}=0.06531(60)$ fm, $a_{\beta=5.2}=0.07513(79)$ fm, determined from a fit in the chiral sector \cite{LottiniRFA}, are considered with pion masses in the range $[190\,, 440]~\MeV$. With respect
to the work reported in \cite{BlossierJOL}, we have taken the bare strange quark masses at $\beta=5.2$ from \cite{DellaMorteDYU} and we have tuned the charm quark mass on those
ensembles by imposing $aM_P(a, \kappa_{\rm sea},\kappa_c, \kappa_s)=am^{\rm physical}_{D_s}$. The values we find for $\kappa_c$ are close to what is quoted in \cite{DellaMorteDYU} where the tuning was realised thanks to a constraint on cut-off effect magnitude for the ratio of PCAC masses $m^{\rm PCAC}_c/m^{\rm PCAC}_s$. We have used the same procedure as in \cite{BlossierJOL} to compute the statistical error at finite $a$ and in the continuum limit, to compute stochastic all-to-all propagators and to reduce the contamination by excited states on 2-pt correlators by solving a $4 \times 4$ Generalized Eigenvalue Problem (GEVP) with one local and 3 Gaussian smeared interpolating fields. In our application of the step scaling in mass strategy we have chosen $K=6$ steps. 
Similarly to \cite{BlossierJOL} we have extracted
the relevant matrix elements from projected correlators along the fixed ground state generalized eigenvectors $v^{(1)}_P(V)(t_{\rm fix},t_0)$.
$(t_{\rm fix}, t_0)$ take the values $(4a, 3a)$ at $\beta=5.2$, $(4a, 3a)$ at $\beta=5.3$ and $(6a, 5a)$ at $\beta=5.5$. We collect in Tables \ref{tabmpcacA5} -- \ref{tabratO7} of the Appendix the whole set of raw data we need in our analysis, i.e. ratios of pseudoscalar and vector heavy-strange mesons masses for 2 subsequent heavy bare quark masses, ratios of pseudoscalar and vector heavy-strange meson decay constants for 2 subsequent heavy bare quark masses and PCAC quark masses $m_{hs}$ defined by:
\begin{equation}
m_{hs}^{\rm PCAC}=\frac{\frac{\partial_0 +\partial^*_0}{2} C_{A_0P}(t) -a c_A \partial_0 \partial^*_0 C_{PP}(t)}{2C_{PP}(t)},
\end{equation}
where $\frac{\partial_0 +\partial^*_0}{2} f(t) = \frac{f(t+a)-f(t-a)}{2a}$, $C_{A_0P}$ and $C_{PP}$ are axial-pseudoscalar and pseudoscalar-pseudoscalar 2-pt correlation functions of the heavy-strange meson defined by the bare quark masses $(\kappa_h, \kappa_s)$ and $c_A$ is the improvement coefficient of the axial bilinear of Wilson-Clover fermions determined in \cite{DellaMorteAQE}.\\
We show in Figure \ref{fig:platmas} the effective mass for two heavy-strange mesons, one at the first step scaling in mass and the other at the next to last step.
For very heavy quarks the signal deteriorates quickly. It explains why we fixed shorter interval ranges to extract the hadronic properties, as indicated in Tables \ref{tabmpcacA5} -- \ref{tabratO7}.

\section{Analysis and discussion\label{sec4}}

\subsection{Extraction of $f_{B_s}$}

\begin{figure}[t]
\begin{tabular}{c}
\begin{tabular}{ccc}
\includegraphics*[width=5cm, height=5cm]{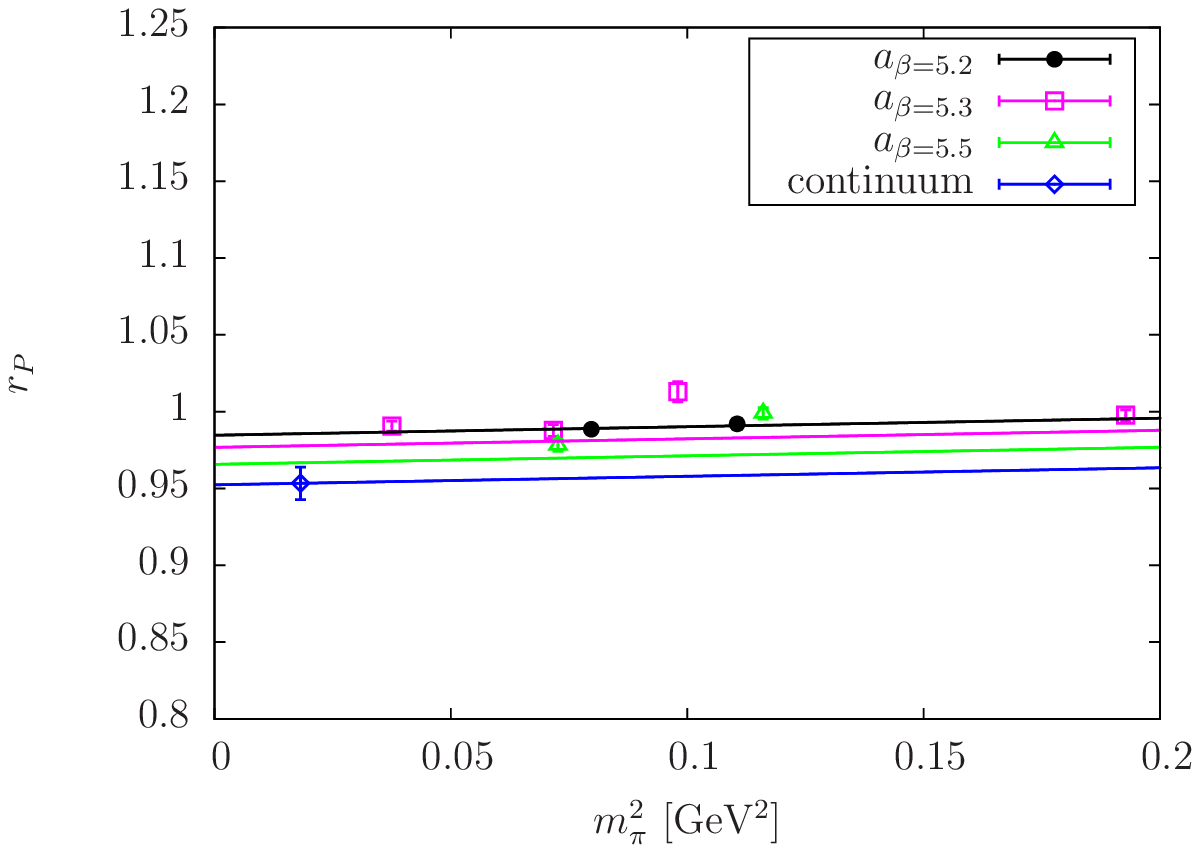}
&
\includegraphics*[width=5cm, height=5cm]{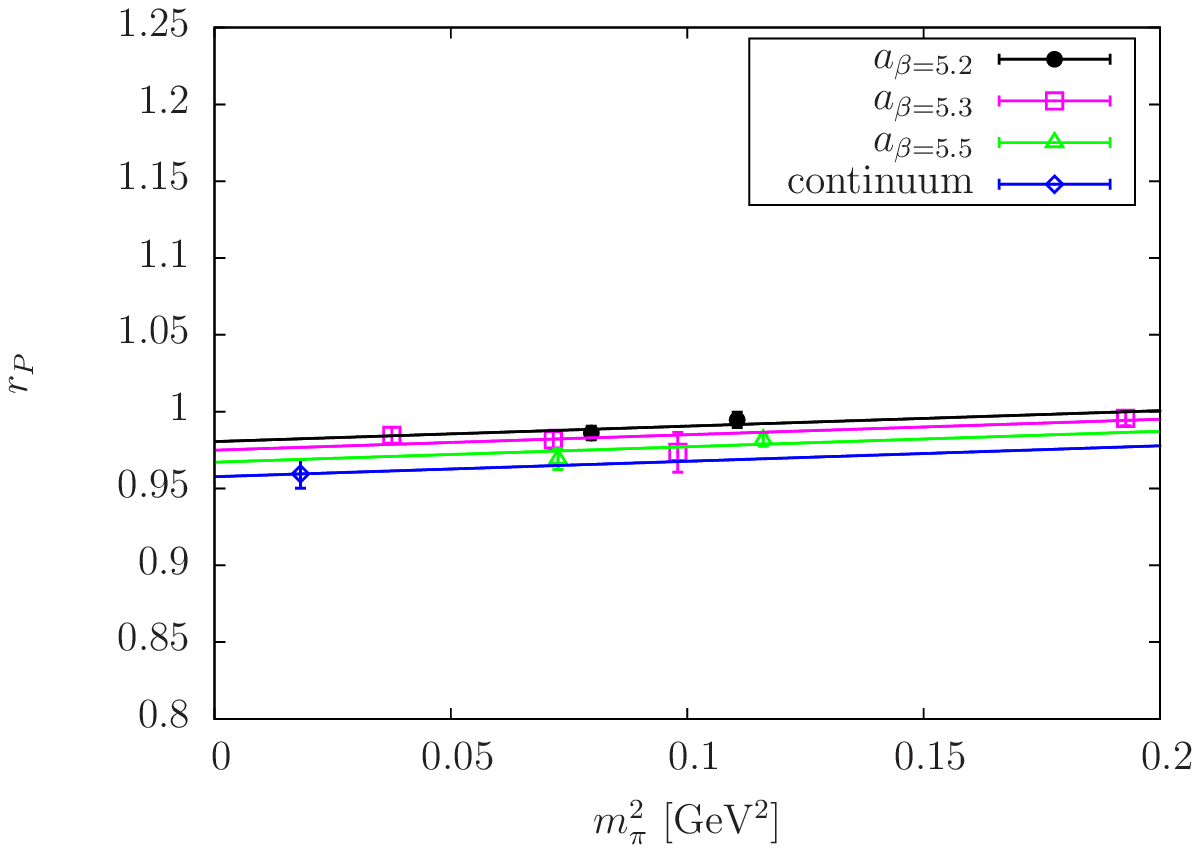}
&
\includegraphics*[width=5cm, height=5cm]{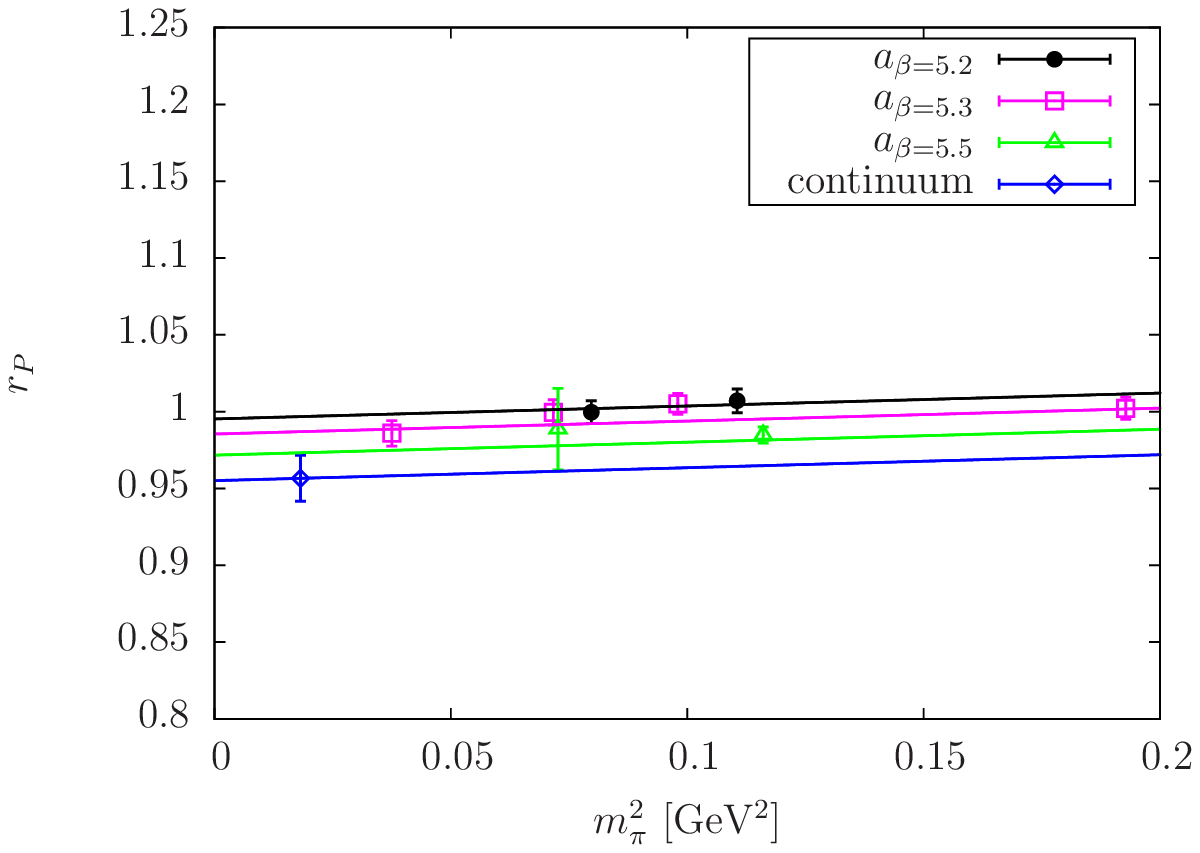}\\
(a)&(b)&(c)\\
\end{tabular}
\\
\\
\begin{tabular}{cc}
\includegraphics*[width=5cm, height=5cm]{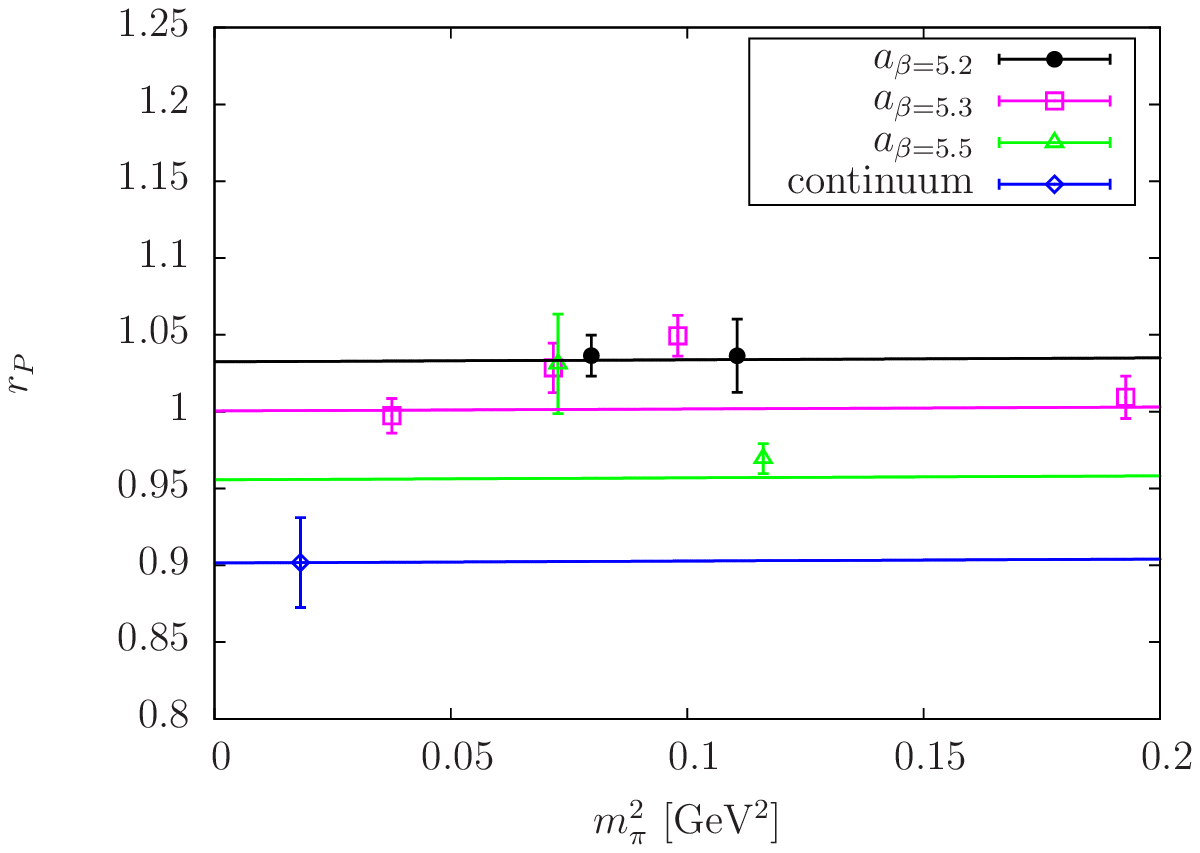}
&
\includegraphics*[width=5cm, height=5cm]{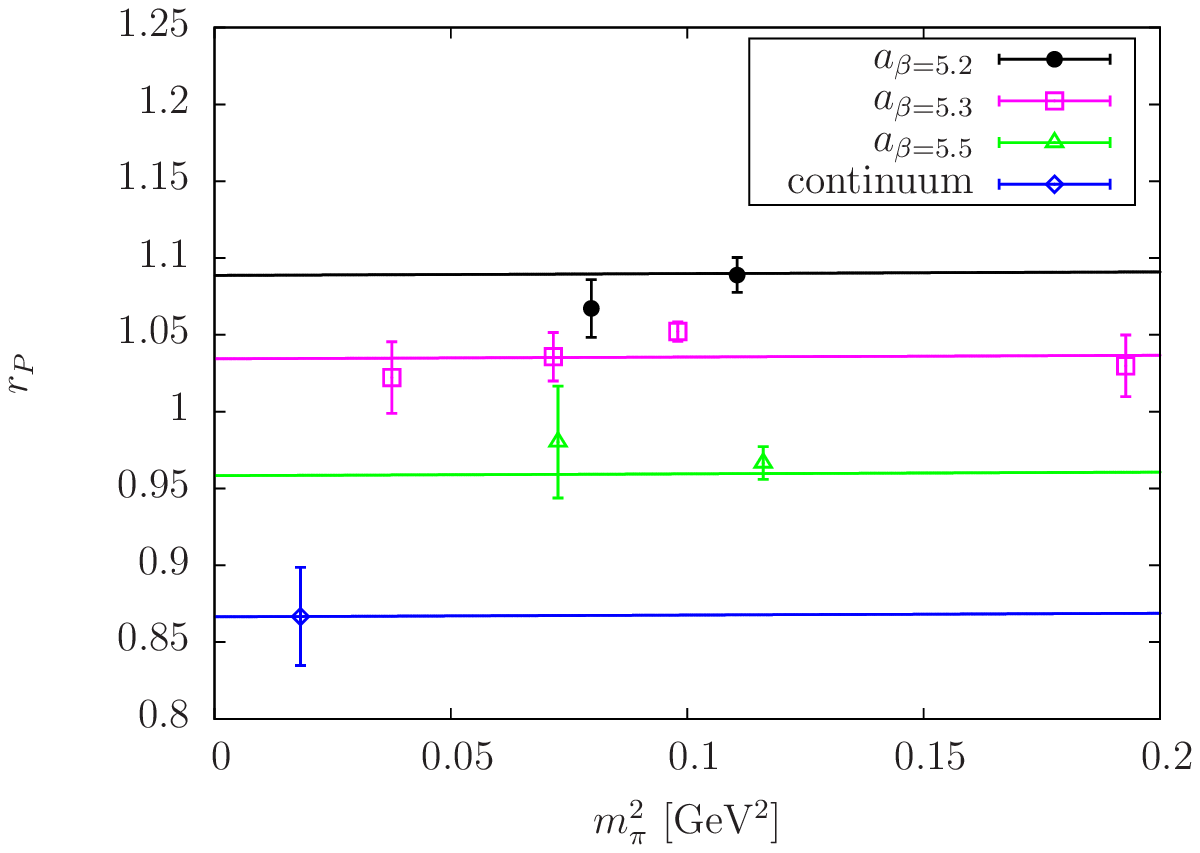}\\
(d)&(e)\\
\end{tabular}
\\
\end{tabular}
\caption{\label{figrP} Continuum and chiral extrapolation of $r_P$ at the heavy masses $\lambda m_{D_s}$ (a), $\cdots$, $\lambda^5 m_{D_s}$ (e).}
\end{figure}

\begin{figure}[t]
\begin{center}
\includegraphics*[width=9cm, height=5cm]{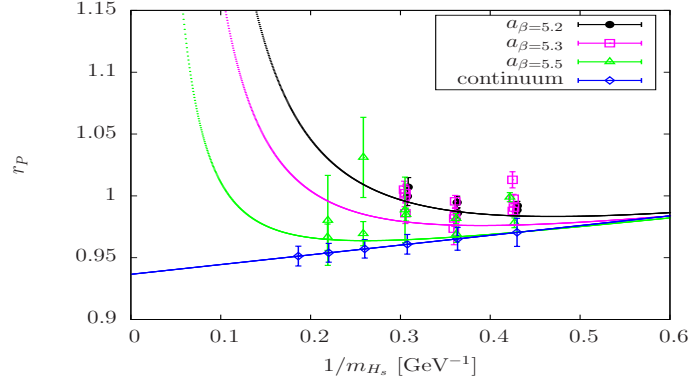}
\end{center}
\caption{\label{figfHsmHsstarfHsstar_restr} Extrapolation at the physical point of $r_P$. The curves correspond to extrapolations at $m_\pi = m^{\rm physical}_\pi$ and $r_{\rm mistuning}-1=0$.}
\end{figure}

We have performed extrapolations to the physical point by doing a global fit analysis. However, in a preparatory stage, we restrict our analysis to a given step in heavy mass $i$ and study the pion mass and the cut-off dependence of $r_P(i)$. Here we ignore the mistuning effects because our goal is to determine how large are the discretisation effects. We show in Figure \ref{figrP} the 
extrapolation in $a^2$ and $m^2_\pi$ of $r_P(i)$.
We observe very big cut-off effects for the $4^{\rm th}$ ratio (10\%) and the $5^{\rm th}$ ratio (17\%). Hence we are not quite confident in using the ensembles at $a=0.075$ fm and, most probably, those at $a=0.065$ fm as well, at the final stage of our analysis. 
\noindent
That is why we prefer, in the combined fit analysis, to exclude the data at the fourth and the fifth heavy quark mass 
at the lattice spacings 0.075 fm and 0.065 fm, Then, we have used the following fit ansatz:
\begin{eqnarray}
\nonumber
r_P(a, m^2_\pi, M_{H_s}(i))&=&1+r_{P_0}+r_{P_1} m^2_\pi + r_{P_2}(r_{\rm mistune}(i) - 1) + r_{P_3}/M_{H_s}(i)\\ 
\label{eqfitratios}&+& r_{P_4} (a/a_{\beta=5.3})^2
+r_{P_5}(a/a_{\beta=5.3})^2 (a M_{H_s}(i))^2,
\end{eqnarray}
with $r_{\rm mistune}(i)=\frac{M_{H_s}(i)}{\lambda^i M_{H_s}(0)}$. We collect in Table \ref{tabparafitextrapoHQET} the fit parameters and $\chi^2/d.o.f.$\footnote{In lattice QCD data analysis, a $\chi^2$/d.o.f. of the order 1 means that the proposed model to describe them is acceptable. In the discussion we have paid more attention to the stability of fit parameters when more terms are added in the fit formula, as well as their compatibility with 0 or not}.
 \begin{table}[t]
 \begin{center}
 \begin{tabular}{|c|c|c|c|c|c|c|}
  \hline
$r_{P_0}$&$r_{P_1}$[${\rm GeV^{-2}}$]&$r_{P_2}$&$r_{P_3}$[GeV]&$r_{P_4}$&$r_{P_{5}}$&$\chi^2/{\rm d.o.f.}$\\
\hline
 -0.06(1)&0.08(3)& 1.1(3)& 0.08(5)&-0.006(3)& 0.021(9)& 1.4\\
  \hline
  \end{tabular}
  \end{center}
   \caption{\label{tabparafitextrapoHQET} Fit parameters of $r_P$ and its $\chi^2/{\rm d.o.f.}$.}
 \end{table}
We show in Figure \ref{figfHsmHsstarfHsstar_restr} the dependence of $r_P$ on $1/M_{H_{s}}$ found with the fit formula (\ref{eqfitratios}). 
We retrieve the parametrically large cut-off effects $\propto (a M_{H_s})^2$ on $r_P$, justifying our decision to exclude some data of our 2 coarsest lattices in the analysis. It is reassuring that the pion mass dependence is found to be of the order of a few \% and that it is numerically a sub-leading effect: by construction of $r_P$, pion mass effects are expected to vanish. Our way to define $r_P$ is such that, according to the HQET scaling law telling that $\lim_{M_P \to \infty} f_P \sqrt{M_P} = C_{\rm ste}$, $r_P$ should tend to 
$\frac{1}{\sqrt{\lambda}}$. With our value of $\lambda=1.18$, the limit is expected to be 0.92, in excellent agreement with our fit parameter $1+r_{P_0}=0.94(2)$.
Then, in the continuum, we interpolate $r_P$ at the 6 points $m_{D_s} \lambda^{i+1}$ to get a set of 6 ratios $r'_P(i)$:
\begin{eqnarray}
\nonumber
r_P(i) &=&1+r_{P_0}+r_{P_1} m^{2\, {\rm physical}}_\pi  + \frac{r_{P_3}}{\lambda^{i+1} m_{D_s}},\\
r'_P(i)&=& \frac{C^{\rm stat}_A(\mu_f,m_{D_s}\lambda^{i+1})}
{C^{\rm stat}_A(\mu_f,m_{D_s}\lambda^i)} r_P(i),
\end{eqnarray}
We recall that $r'_P(i)$ is independent of the renormalization scale $\mu_f$  and that we have $r'_P(i)=\frac{f_{H_s}(i+1)}{f_{H_s}(i)}$. We collect in Table \ref{tabfitinterpol} values of $r'_P$ at the reference points $\frac{1}{\lambda^{i+1} m_{D_s}}$.
\begin{table}[t]
 \begin{center}
 \begin{tabular}{|c|c|c|c|c|c|c|}
 \hline
$1/(m_{D_s} \lambda^{i+1})$ [${\rm GeV}^{-1}$]&0.4298&0.3637&0.3077&0.2603&0.2202&0.1863\\
\hline
 $r'_P$&0.9945(116)&0.9863(94)&0.9795(81)&0.9739(76)&0.9692(77)&0.9653(81)\\
\hline       
\end{tabular}
\end{center}
\caption{\label{tabfitinterpol}Ratio $r'_P$ at the reference points 
$\frac{1}{\lambda^{i+1} m_{D_s}}$.}
\end{table}
The last step is straightforward: $f_{B_s}/f_{D_s}$ is obtained by a series of products:
\begin{equation}
\frac{f_{B_s}}{f_{D_s}}= \prod_{i=0}^{5} r'_P(i). 
\end{equation}
We get
\begin{equation}
\frac{f_{B_s}}{f_{D_s}}=0.88(4).
\end{equation}
The effect on the statistical error of the correlation among the different terms of the product of $r'_P(i)$ is taken into account by the mean of computing the errors described in \cite{BlossierJOL}.

To address the systematic error, we have performed two other fits:\\
-- fit(A), adding to (\ref{eqfitratios}) a ``next to leading order" chiral contribution in $m^2_\pi \ln (m^2_\pi)$\\
-- fit(B), adding to (\ref{eqfitratios}) a contribution in $1/m^2_{H_s}(i)$ to count for a higher order in the heavy quark expansion\\
-- fit(C): fit (\ref{eqfitratios}) but using the matching coefficient $C^{\rm stat}_A$ at NLO\\

Other fits with extra terms in $(aM_{H_s})^2$ or in $a^3$ give non reliable results. We collect the corresponding fit parameters and $
\chi^2/d.o.f.$ in Table \ref{tabfitfHstry} and we get
\beq
\frac{f_{B_s}}{f_{D_s}}=0.87(6)\, (A), \quad \frac{f_{B_s}}{f_{D_s}}=0.88(4)\, (B), \quad \frac{f_{B_s}}{f_{D_s}}=0.87(4)\, (C).
\eeq
\begin{table}[t]
 \begin{center}
 \begin{tabular}{|c|c|c|c|c|c|c|c|c|}
 \cline{2-9}
\multicolumn{1}{l|}{}&$r_{P_0}$&$r_{P_1}$[${\rm GeV}^{-2}$]&$r_{P_2}$&$r_{P_3}$[GeV]&$r_{P_4}$&$r_{P_5}$&$r_{P_{{\rm fit}_i}}$
&$\frac{\chi^2}{{\rm dof}}$\\
\hline
A&-0.07(2)&0.4(1.7)& 1.1(3)& 0.08(5)&-0.01(1)& 0.023(9)&-0.03(13) ${\rm GeV}^{-2}$& 1.5\\
B& -0.05(6)&0.08(3)& 1.1(3)&-0.03(32)&-0.01(1)& 0.022(8)& 0.2(4) ${\rm GeV}^2$& 1.5\\
C&-0.07(2)&0.08(3)& 1.1(3)& 0.09(5)&-0.01(2)& 0.021(9)&-& 1.4\\
\hline
\end{tabular}
 \end{center}
  \caption{\label{tabfitfHstry} Fit parameters of $r_P$ and the respective $\chi^2/{\rm d.o.f.}$ for the fits (A), (B) and (C).
  (A) corresponds to adding an NLO term in $m^2_pi \ln m^2_pi$ to eq.(\ref{eqfitratios}), (B) corresponds to adding a term in $1/m^2_{H_s}$ to eq.(\ref{eqfitratios}) while (C) is using the expression (\ref{eqfitratios}) and the formulae at NLO of the matching coefficient $C^{\rm stat}_A$ to get $r_P$,} 
  \end{table}
A fourth source of systematics can be included by propagating the uncertainty on raw data if we change $t_{\rm min} \to t_{\rm min}+2a$ to extract plateaus. In this case we get
the following result:
\begin{equation}
\frac{f_{B_s}}{f_{D_s}}=0.89(5).
\end{equation}
We collect the corresponding fit parameters in Table \ref{tabfitsysterr}.
\begin{table}[t]
 \begin{center}
 \begin{tabular}{|c|c|c|c|c|c|c|}
 \hline
$r_{P_0}$&$r_{P_1}$[${\rm GeV}^{-2}$]&$r_{P_2}$&$r_{P_3}$[GeV]&$r_{P_4}$&$r_{P_5}$&$\chi^2/{\rm d.o.f.}$\\
\hline
-0.06(2)&0.07(3)& 1.0(3)& 0.07(5)&-0.01(2)& 0.021(9)& 0.8\\
\hline
\end{tabular}
 \end{center}
    \caption{\label{tabfitsysterr} Fit parameters of $r_P$ and its $\chi^2/{\rm d.o.f.}$ when systermatic errors on
   raw data are propagated in the analysis.}
  \end{table}
Adding together the different sources of systematics, we obtain
 \begin{equation}
  \frac{f_{B_s}}{f_{D_s}}=0.88(4)(^{+1}_{-2}). 
 \end{equation}
 where the first error is statistical and the second error counts for the systematic error.

Concerning the $D_s$ meson decay constant, an update of the analysis reported in \cite{BlossierJOL}, now that we have the additional coarsest ensembles A5 and B6, meaning a third lattice spacing at our disposal, 
gives  
\begin{equation}
f_{D_s}=244(4)(2)\, {\rm MeV}, \quad f_{D^*_s}=268(4)(2)\, {\rm MeV}, \quad f_{D^*_s}/f_{D_s}=1.10(2),
\end{equation}
 where the first error is statistical and the second error comes from the uncertainty on the lattice spacings. As in the previous analysis, a next to leading order contribution to the chiral fir destabilises the fit with chiral fit parameters compatible with zero.

Then we get 
\begin{equation}
f_{B_s}=215(10)(2)(^{+2}_{-5})\, {\rm MeV},
\end{equation}
 where the first error is the statistical error, the second one counts for the systermatic error on $f_{D_s}$ while the third error corresponds to the systematic error on $f_{B_s}/f_{D_s}$. 
 
 FLAG has recently made an update collection of lattice estimates of $f_{B_s}$ \cite{AokiCCA}. Our estimate of $f_{B_s}$ using the step scaling in mass strategy is compatible with the value obtained by the ALPHA Collaboration $f_{B_s}=224(14)$ MeV \cite{BernardoniFVA} by a computation, performed over almost the same CLS ensembles as in this paper, of hadronic matrix elements in the framework of HQET with a non-perturbative matching of the HQET parameters with QCD. It is 2$\sigma$ lower than the result reported by the ETM Collaboration \cite{CarrascoZTA} with $N_f=2$ twisted-mass fermions defined at maximal twist.  The fact that we cannot constrain the static limit of the ratio $r_P$ to be equal to 1, due to mistuning effects of the heavy quark mass, explains a part of that discrepancy. The second source of discrepancy is the presence of large $a^2 (a M_{H_s})^2$ cut-off effects in our data while they are numerically absent in ETMC data. Having to take them into account necessarily increases the uncertainty in extrapolation to the continuum limit because more parameters are required to described the data.

\subsection{Extraction of $f_{B^*_s}/f_{B_s}$}

\begin{figure}[t]
\begin{center}
\begin{tabular}{cc}
\includegraphics*[width=7cm, height=5cm]{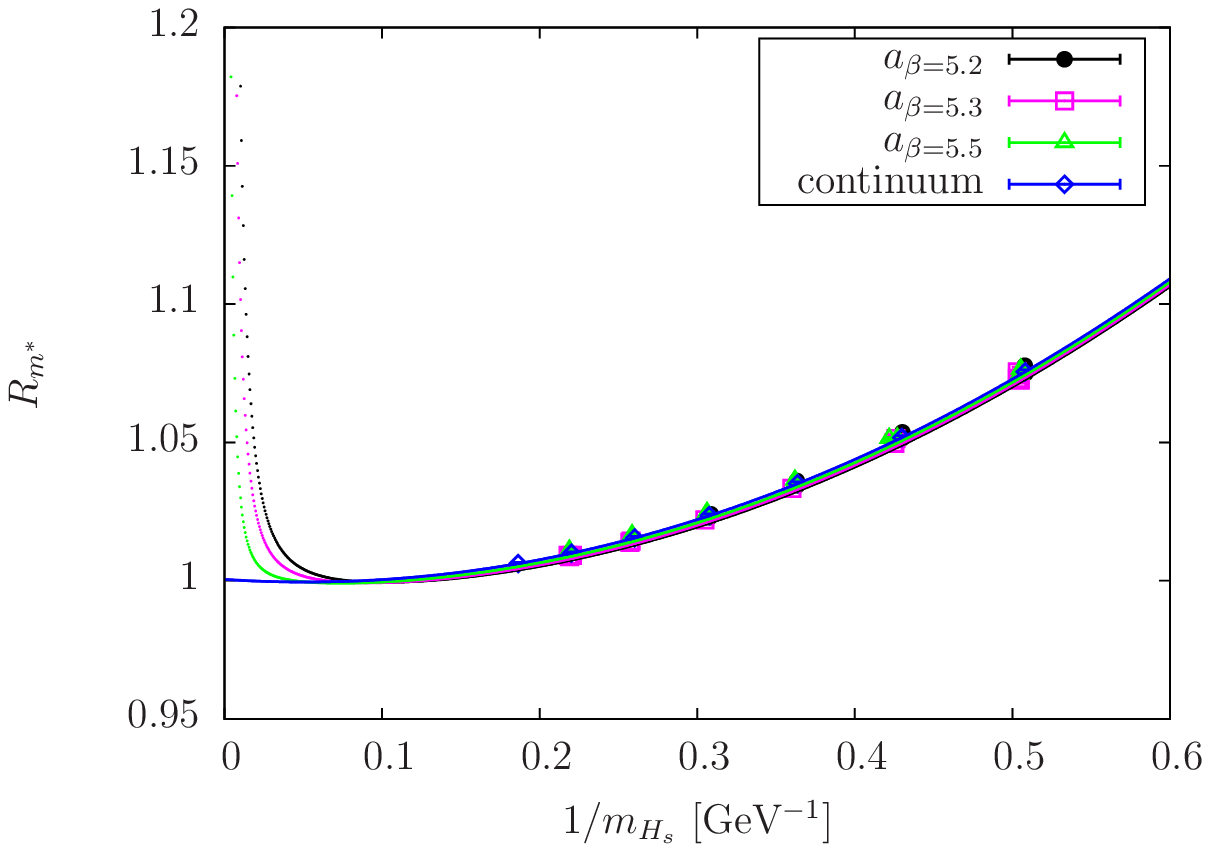}
&
\includegraphics*[width=7cm, height=5cm]{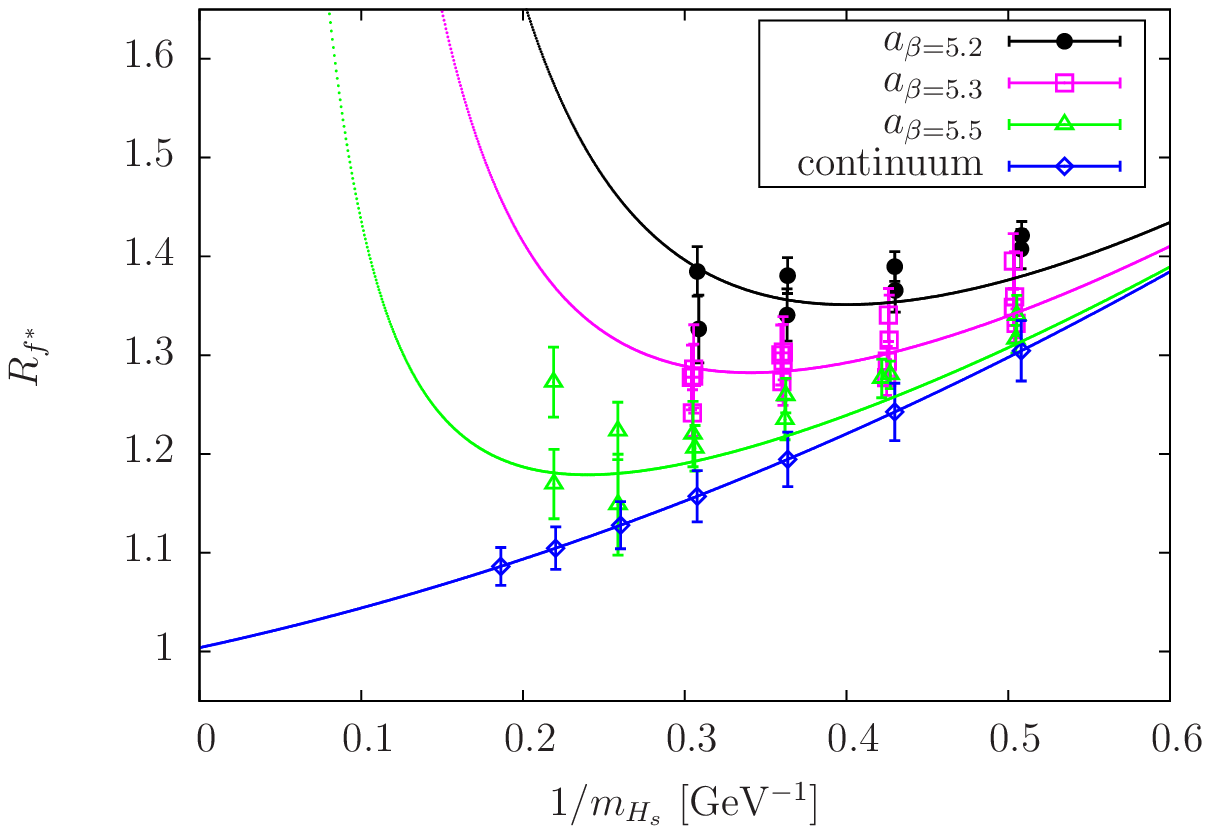}\\
(a)&(b)\\
\end{tabular}
\end{center}
\caption{\label{figBsstarBs}Extrapolations at the physical point of $R_{m^*}$ (a) and $R_{f^*}$ (b). The curves correspond to extrapolations at $m_\pi = m^{\rm physical}_\pi$.}
\end{figure}

To extract $f_{B^*_s}/f_{B_s}$ we have performed an alternative analysis to the one discussed in the previous subsection. We have examined the ratios 
\begin{eqnarray}
\nonumber
R_{m^*}&=&\frac{M_V(a,\kappa_{\rm sea},\kappa_h,\kappa_s)}{M_P(a,\kappa_{\rm sea},\kappa_h,\kappa_s)}\equiv \frac{M_{H^*_s}}{M_{H_s}},\\ 
\nonumber
R'_{f^*}&=&\frac{f_V(a,\kappa_{\rm sea},\kappa_h,\kappa_s)}{f_P(a,\kappa_{\rm sea},\kappa_h,\kappa_s)}\equiv \frac{f_{H^*_s}}{f_{H_s}}\\
R_{f^*}(a, m^2_\pi, M_{H_s})&\equiv&\frac{C^{\rm stat}_V(\mu_f,M_{H_s})}{C^{\rm stat}_A(\mu_f,M_{H_s})} R'_{f^*}.
\end{eqnarray}
As the HQET anomalous dimension of the axial and vector static-light operator are the same, applying the renormalization group equation makes $R_{f^*}$ independent of the renormalization scale $\mu_f$.
To extrapolate to the physical point we have used the following fit ansatz:
\beq\label{eqrmstar}
R_{m^*}(a, m^2_\pi, M_{H_s})=1+r_{m^*_0} m^2_\pi + r_{m^*_1}/M_{H_s} + r_{m^*_2} (a/a_{\beta=5.3})^2 + r_{m^*_3} /M^2_{H_s} + r_{m^*_4} (a/a_{\beta=5.3})^2 (a M_{H_s})^2,
\eeq
\beq
R_{f^*}(a, m^2_\pi, M_{H_s})=1+r_{f^*_0} m^2_\pi + r_{f^*_1}/M_{H_s} + r_{f^*_2} (a/a_{\beta=5.3})^2 
+r_{f^*_3} /M^2_{H_s}+r_{f^*_4} (a/a_{\beta=5.3})^2 (a M_{H_s})^2.
\label{eqrfstar}
\eeq
We can impose the static limit constraint $\lim_{M_{H_s} \to \infty} R_{m^*} = \lim_{M_{H_s} \to \infty} R_{f^*}=1$ because those ratios are free of heavy quark mistuning effect. We collect in Table \ref{tabfitmBsstarmBsalter} the corresponding fit parameters and we obtain
\beq\nonumber
\frac{m_{B^*_s}}{m_{B_s}}=1.0061(4),\quad \left(\frac{m_{B^*_s}}{m_{B_s}}\right)^{\rm exp}=1.0091, \quad \frac{f_{B^*_s}}{f_{B_s}}=1.02(2).
\eeq
\begin{table}[t]
 \begin{center}
 \begin{tabular}{|c|c|c|c|c|c|c|}
 \cline{2-7}
\multicolumn{1}{l|}{}&$r_{X_0}$[${\rm GeV}^{-2}$]&$r_{X_1}$[GeV]&$r_{X_2}$&$r_{X_3}$[${\rm GeV}^2$]&$r_{X_4}$&$\chi^2/{\rm d.o.f.}$\\
\hline
$X\equiv m^*$&0.026(3)&-0.007(3)&-0.002(1)& 0.364(4)& 0.0001(1)& 1.7\\
\hline
$X\equiv f^*$&0.2(2)& 0.4(2)&-0.01(4)& 0.5(2)&0.12(3)&1.2\\
\hline
\end{tabular}
 \end{center}
  \caption{\label{tabfitmBsstarmBsalter} Fit parameters of $R_{m^*}$ and $R_{f^*}$ and their respective $\chi^2/{\rm d.o.f.}$ .}
  \end{table}
We show in Figure \ref{figBsstarBs} the extrapolation to the physical point of $R_{m^*}$ and $R_{f^*}$.

To estimate the systermatic error, we have performed fits (A') and (B'), that read\\
-- fit(A'): add to (\ref{eqrmstar}) and (\ref{eqrfstar}) a contribution in $m^2_\pi \ln (m^2_\pi)$\\
-- fit(B'): add to (\ref{eqrmstar}) a contribution in $1/m^3_{H_s}$\\
-- fit(C'): fit (\ref{eqrfstar}) but using matching coefficients $C^{\rm stat}_A$ and $C^{\rm stat}_V$  at NLO\\

Other terms in the fit lead to unstable and reliable results. We collect the fit parameters in 
Tables \ref{tabfitsystmBsstarmBs} and \ref{tabfitsystfBsstarfBs} and we obtain:
\beq\nonumber
\frac{m_{B^*_s}}{m_{B_s}}=1.0059(7), \quad \frac{f_{B^*_s}}{f_{B_s}}=1.02(2) \quad (A'),
\eeq
\beq\nonumber
 \frac{m_{B^*_s}}{m_{B_s}}=1.0058(7) \quad (B'), \quad \frac{f_{B^*_s}}{f_{B_s}}=1.03(2) \quad (C'). 
\eeq
 \begin{table}[t]
 \begin{center}
 \begin{tabular}{|c|c|c|c|c|c|c|c|}
 \cline{2-8}
\multicolumn{1}{l|}{}&$r_{m^*_0}$[${\rm GeV}^{-2}$]&$r_{m^*_1}$[GeV]&$r_{m^*_2}$&$r_{m^*_3}$[${\rm GeV}^2$]&$r_{m^*_4}$&$r_{m^*_{{\rm fit}_i}}$&$\frac{\chi^2}{{\rm dof}}$\\
\hline
A'&0.1(2)&-0.042(9)&-0.001(1)& 0.37(1)&0.0004(10)&0.01(1) ${\rm GeV}^{-2}$&1.7\\
B'&0.027(4)&-0.053(5)&-0.005(1)& 0.46(2)& 0.0020(1)&-0.11(2) ${\rm GeV}^3$&1.4\\
\hline
\end{tabular}
 \end{center}
  \caption{\label{tabfitsystmBsstarmBs} Fit parameters of $R_{m^*}$ and its $\chi^2/{\rm d.o.f.}$ for the fits (A') and (B').
  (A') corresponds to adding an NLO term in $m^2_pi \ln m^2_pi$ to eq.(\ref{eqrmstar}) and (B') corresponds to adding a term in $1/m^3_{H_s}$ to eq.(\ref{eqrmstar}),}
  \end{table}
 \begin{table}[t]
 \begin{center}
 \begin{tabular}{|c|c|c|c|c|c|c|c|}
 \cline{2-8}
\multicolumn{1}{l|}{}&$r_{f^*_0}$[${\rm GeV}^{-2}$]&$r_{f^*_1}$[GeV]&$r_{f^*_2}$&$r_{f^*_3}$[${\rm GeV}^2$]&$r_{f^*_4}$&$r_{f^*_{{\rm fit}}}$&$\frac{\chi^2}{{\rm dof}}$\\
\hline
 A'&2(10)&0.3(4)&-0.01(4)& 0.6(5)&0.12(3)&-0.2(8) ${\rm GeV}^{-2}$&1.3\\
C'&0.2(2)&0.3(1)&-0.01(4)& 0.4(2)&0.12(3)&-&1.2\\
\hline
\end{tabular}
 \end{center}
  \caption{\label{tabfitsystfBsstarfBs} Fit parameters of $R_{f^*}$ and its $\chi^2/{\rm d.o.f.}$ for the fits (A') and (C').
  (A') corresponds to adding an NLO term in $m^2_pi \ln m^2_pi$ to eq.(\ref{eqrfstar}) and (C') is using the expression (\ref{eqrfstar}) and the formulae at NLO of the matching coefficient $C^{\rm stat}_A$ and $C^{\rm stat}_V$ to get $R_{f^*}$,}
  \end{table}
As for $f_{B_s}/f_{D_s}$, we have counted for the systematic error coming from the change $t_{\rm min} \to t_{\rm min}+2a$ in the plateaus extraction. We get
the following results:
\begin{equation}
\frac{m_{B^*_s}}{m_{B_s}}=1.0066(7), \quad \frac{f_{B^*_s}}{f_{B_s}}=1.03(2).
\end{equation}
Fit parameters are collected in Table \ref{tabfitsysterrmBsstarfBsstar}.
\begin{table}[ht]
 \begin{center}
 \begin{tabular}{|c|c|c|c|c|c|c|}
 \cline{2-7}
\multicolumn{1}{l|}{}&$r_{X_0}$[${\rm GeV}^{-2}$]&$r_{X_1}$[GeV]&$r_{X_2}$&$r_{X_3}$[${\rm GeV}^2$]&$r_{X_4}$&$\chi^2/{\rm d.o.f.}$\\
\hline
$X\equiv m^*$&0.02(1) &-0.03(6)&-0.003(1)& 0.364(7)&0.001(1)&0.4\\
$X\equiv f^*$&0.3(2)&0.4(2)&-0.03(5)&0.4(2)&0.13(4)&  0.7\\
\hline
\end{tabular}
 \end{center}
  \caption{\label{tabfitsysterrmBsstarfBsstar} Fit parameters of $ R_{m^*}$ and $R_{f^*}$ and their respective $\chi^2/{\rm d.o.f.}$ when systermatic errors on
  raw data are included.}
  \end{table}

 Eventually, we quote
 \begin{equation}
  \frac{m_{B^*_s}}{m_{B_s}}=1.0061(4)(5), \quad \frac{f_{B^*_s}}{f_{B_s}}=1.02(2)(^{+2}_{-0}),
 \end{equation}
 where the first error is statistical and the second error counts for the systematic error estimated by the fits (A') and (B') and contamination from
 excited states. There is no reason why the ratio $ \frac{m_{B^*_s}}{m_{B_s}}$ should correspond to the experimental ratio $\left(\frac{m_{B^*_s}}{m_{B_s}}\right)^{\rm exp}=1.0091$ because, in our analysis, the strange quark is quenched. Still, we find a ratio 1$\sigma$ lower than in \cite{BernardoniNQA} (1.0070(6)) where the computation was done in the framework of HQET expanded at ${\cal O}(1/m_b)$.
 
 \subsection{Comment}
 
\begin{figure}[t]
\begin{center}
\begin{tabular}{cc}
\includegraphics*[width=7cm, height=5cm]{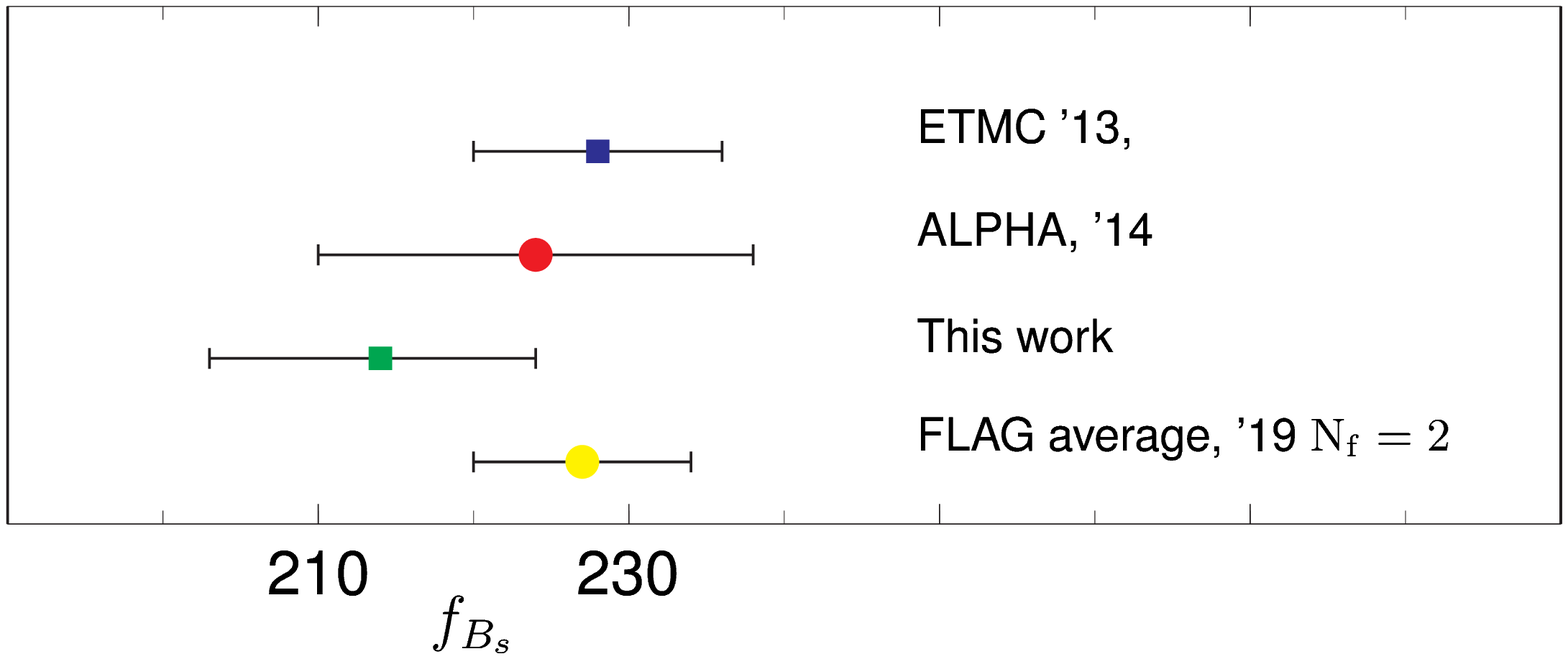}
&
\includegraphics*[width=7cm, height=5cm]{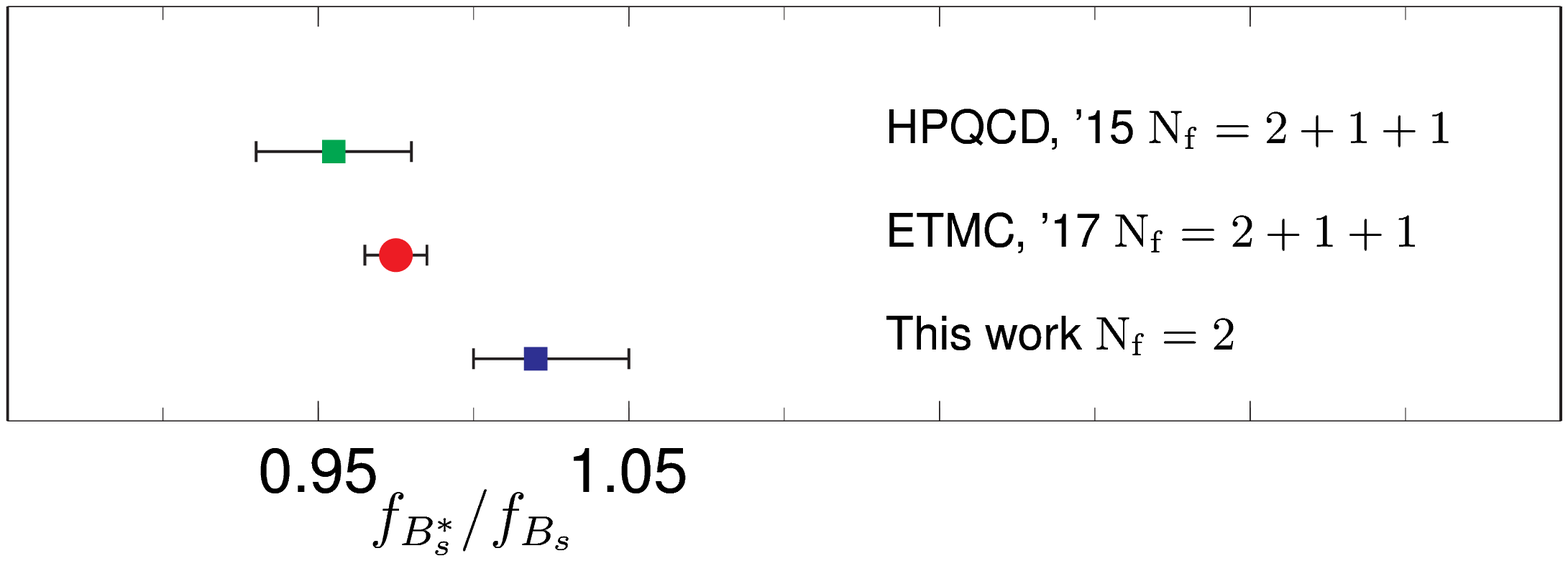}\\
(a)&(b)\\
\end{tabular}
\end{center}
\caption{\label{figfBs} Lattice estimates of $f_{B_s}$ at ${\rm N_f}=2$, with the FLAG average of 2016 (a) and of $f_{B^*_s}/f_{B_s}$ (b).}
\end{figure}

We collect in Figure \ref{figfBs} the lattice QCD estimates of $f_{B_s}$ at $N_f=2$ \cite{CarrascoZTA}, \cite{BernardoniFVA}, with the corresponding FLAG average \cite{AokiCCA} and those of $f_{B^*_s}/f_{B_s}$ \cite{ColquhounOHA}. \cite{LubiczASP}.
Of course the fact that we get $f_{B^*_s}/f_{B_s}>1$ while the 2 other lattice QCD results read $f_{B^*_s}/f_{B_s}<1$ is puzzling. However, 
a computation performed by the ETM Collaboration with ${\rm N_f}=2$ dynamical quarks indicated the hierarchy $f_{B^*}/f_B>1$, $f_{B^*}/f_B=1.050(16)$ 
\cite{BecirevicKAA}. So, a plausible explanation for the observed tension is the effect of the quenching of the strange quark in the spin-breaking contribution of the
heavy quark symmetry to the ratio $f_{B^*_s}/f_{B_s}$. It might be of the same order of magnitude as in $f_{D^*_s}f_{D_s}$ but with a more important
qualitative impact because we examine a region of paramaters closer to the symmetric point $f_{H^*}/f_H=1$. In that respect, studies of this ratio with ${\rm N_f}=2+1$
ensembles are welcome.

\section{Conclusion\label{sec5}}

In that paper we have reported on a lattice estimate of $f_{B_s}$ and $f_{B^*_s}/f_{B_s}$. The main puropose of the work was testing the step scaling in mass method 
with Wilson-Clover fermions for which the RGI heavy quark mass can not be used yet as a physical parameter of the heavy quark expansion. Indeed, severe negative $O(am)$ cut-effects
need to be balanced by still unknown $O(am)^2$ improvement terms to define safely the RGI mass. Instead, we have chosen the (inverse of) the heavy-strange meson
mass as the expansion parameter. We obtain a quite low result for $f_{B_s}$ compared to other lattice QCD estimates at ${\rm N_f}=2$, though it is compatible with the one got using the
same set of gauge ensembles as here but with a complete different approach to simulate the heavy quark. We have found the hierarchy $f_{B^*_s}/f_{B_s}>1$, indicating a positive correction to the symmetric point when the strange quark 
is quenched. A look at the literature leads to the conclusion that this correction becomes negative when the strange quark is taken into account in the sea. The next step of our
program is the investigation of the form factors associated to $B_s \to D^{(*)}_s l \nu$ using CLS ${\rm N_f}=2$ ensembles, applying the step scaling in mass method.

\section*{Acknowledgement}

This work was granted access to the HPC resources of CINES and IDRIS under the allocations 2017-x2016056808 
and 2018-A0010506808 made by GENCI. It is supported by Agence Nationale de la Recherche under the contract ANR-17-CE31-0019. Authors are grateful to Olivier P\`ene and Vincent Mor\'enas for useful discussions and the colleagues of the CLS effort for having provided the gauge ensembles used in that work

\section*{Appendix}

In this Appendix we collect all the data (meson masses, PCAC heavy+strange quark masses, meson decay constants), in lattice units, that are used in our analysis. We indicate the time range of the plateaus extraction.

\begin{table}[t]
\begin{center}
\begin{tabular}{|c|c|c|}
\hline
$\kappa_h$&$[t_{\rm min}$ -- $t_{\rm max}]$&$a m^{\rm PCAC}_{hs}$\\
\hline
0.125310&[8--26]& 0.1530(6)\\
 0.121344&[8--26]& 0.2095(7)\\
 0.116040&[8--24]& 0.2910(9)\\
 0.109307&[8--20]& 0.4046(10)\\
 0.100407&[8--18]& 0.5783(13)\\
 0.089289&[8--14]& 0.8484(17)\\
\hline
\end{tabular}
\end{center}
\caption{\label{tabmpcacA5}Average PCAC heavy and strange quark masses for the ensemble A5.}
\end{table}

\begin{table}[t]
\begin{center}
\begin{tabular}{|c|c|c|c|c|c|}
\hline
$\kappa_h$&$[t_{\rm min}$ -- $t_{\rm max}]$&$a M_{H_s}(i)$&$M_{H_s}(i+1)/M_{H_s}(i)$&$a M_{H^*_s}(i)$&$M_{H^*_s}(i+1)/M_{H^*_s}(i)$\\
\hline
 0.125310&[10--24]& 0.7498(8)& -& 0.8081(13)& -\\
 0.121344&[10--24]& 0.8851(8)& 1.1805(4)& 0.9326(13)& 1.1541(4)\\
 0.116040&[10--24]& 1.0482(8)& 1.1843(3)& 1.0860(13)& 1.1645(4)\\
 0.109307&[10--24]& 1.2335(9)& 1.1768(3)& 1.2631(14)& 1.1631(4)\\
 0.100407&[10--20]& 1.4572(11)& 1.1814(5)& 1.4790(17)& 1.1709(4)\\
 0.089289&[10--20]& 1.7202(15)& 1.1804(3)& 1.7348(19)& 1.1730(3)\\
\hline
$\kappa_h$&$[t_{\rm min}$ -- $t_{\rm max}]$&$a f_{H_s}(i)$&$f_{H_s}(i+1)/f_{H_s}(i)$
&$a f_{H^*_s}(i)$&$f_{H^*_s}(i+1)/f_{H^*_s}(i)$\\
\hline
0.125310&[8--24]&0.0911(15)&-&0.1149(12)&-\\
0.121344&[ 8--24]&-& 1.017(4)&-& 0.999(1)\\
 0.116040&[ 8--24]&-& 1.016(5)&-& 1.007(1)\\
 0.109307&[ 8--24]&-& 1.028(8)&-& 1.026(2)\\
 0.100407&[ 8--20]&-& 1.055(24)&-& 1.088(8)\\
 0.089289&[ 8--20]&-& 1.107(12)&-& 1.144(2)\\
\hline
\end{tabular}
\end{center}
\caption{\label{tabratA5} Heavy-strange meson masses and decay constants for the ensemble A5.}
\end{table}

\newpage

\begin{table}[t]
\begin{center}
\begin{tabular}{|c|c|c|}
\hline
$\kappa_h$&$[t_{\rm min}$ -- $t_{\rm max}]$&$a m^{\rm PCAC}_{hs}$\\
\hline
0.125290&[8--38]& 0.1533(5)\\
 0.121313&[8--30]& 0.2105(6)\\
 0.116044&[8--24]& 0.2911(7)\\
 0.109127&[8--22]& 0.4086(8)\\
 0.100172&[8--16]& 0.5871(8)\\
 0.088935&[8--14]& 0.8690(10)\\
\hline
\end{tabular}
\end{center}
\caption{\label{tabmpcacB6} Average PCAC heavy and strange quark masses for the ensemble B6.}
\end{table}

\begin{table}[t]
\begin{center}
\begin{tabular}{|c|c|c|c|c|c|}
\hline
$\kappa_h$&$[t_{\rm min}$ -- $t_{\rm max}]$&$a M_{H_s}(i)$&$M_{H_s}(i+1)/M_{H_s}(i)$&$a M_{H^*_s}(i)$&$M_{H^*_s}(i+1)/M_{H^*_s}(i)$\\
\hline
0.125290&[10--35]& 0.7492(6)&-& 0.8053(11)&-\\
 0.121313&[10--35]& 0.8858(7)& 1.1824(2)&0.9314(10)& 1.1565(4)\\
 0.116044&[10--35]& 1.0473(8)& 1.1823(3)& 1.0835(11)& 1.1633(3)\\
 0.109127&[10--25]& 1.2372(10)& 1.1813(3)& 1.2655(12)& 1.1680(2)\\
 0.100172&[10--20]& 1.4614(11)& 1.1813(2)&1.4830(14)& 1.1719(3)\\
 0.088935&[10--17]& 1.7250(14)& 1.1803(3)& 1.7407(17)& 1.1738(3)\\
\hline
$\kappa_h$&$[t_{\rm min}$ -- $t_{\rm max}]$&$a f_{H_s}(i)$&$f_{H_s}(i+1)/f_{H_s}(i)$
&$a f_{H^*_s}(i)$&$f_{H^*_s}(i+1)/f_{H^*_s}(i)$\\
\hline
0.125290&[10--25]&0.0897(9)&-&0.1142(10)&-\\
  0.121313&[10--25]&-& 1.013(3)&-& 1.003(1)\\
 0.116044&[10--25]&-& 1.007(4)&-& 1.011(1)\\
 0.109127&[10--25]&-& 1.019(8)&-& 1.031(2)\\
 0.100172&[10--22]&-& 1.055(14)&-& 1.081(8)\\
 0.088935&[10--17]&-& 1.085(19)&-& 1.165(9)\\
\hline
\end{tabular}
\end{center}
\caption{\label{tabratB6} Heavy-strange meson masses and decay constantsfor the ensemble B6.}
\end{table}

\newpage

\begin{table}[t]
\begin{center}
\begin{tabular}{|c|c|c|}
\hline
$\kappa_h$&$[t_{\rm min}$ -- $t_{\rm max}]$&$a m_{\rm PCAC}(\kappa_h,\kappa_{\rm s})$\\
\hline
0.127240&[10--25]& 0.1357(9)\\
 0.123874&[10--25]& 0.1833(10)\\
 0.119457&[10--25]& 0.2484(14)\\
 0.113638&[10--25]& 0.3403(17)\\
 0.106031&[10--20]& 0.4744(20)\\
 0.096555&[10--18]& 0.6713(25)\\
\hline
\end{tabular}
\end{center}
\caption{\label{tabmpcacE5} PCAC masses for the ensemble E5}
\end{table}

\begin{table}[t]
\begin{center}
\begin{tabular}{|c|c|c|c|c|c|}
\hline
$\kappa_h$&$[t_{\rm min}$ -- $t_{\rm max}]$&$a M_{H_s}(i)$&$M_{H_s}(i+1)/M_{H_s}(i)$&$a M_{H^*_s}(i)$&$M_{H^*_s}(i+1)/M_{H^*_s}(i)$\\
\hline
0.127240&[10--25]& 0.6579(7)&-& 0.7075(14)&-\\
 0.123874&[10--25]& 0.7770(8)& 1.1810(4)& 0.8169(14)& 1.1546(6)\\
 0.119457&[10--25]& 0.9170(10)& 1.1803(4)& 0.9487(15)& 1.1614(5)\\
 0.113638&[10--25]& 1.0833(13)& 1.1813(3)& 1.1081(17)& 1.1680(4)\\
 0.106031&[10--20]& 1.2826(17)& 1.1840(6)& 1.3010(18)& 1.1741(9)\\
 0.0965545&[10--20]& 1.5117(18)& 1.1786(2)& 1.5248(19)& 1.1720(3)\\
\hline
$\kappa_h$&$[t_{\rm min}$ -- $t_{\rm max}]$&$a f_{H_s}(i)$&$f_{H_s}(i+1)/f_{H_s}(i)$
&$a f_{H^*_s}(i)$&$f_{H^*_s}(i+1)/f_{H^*_s}(i)$\\
\hline
0.127240&[10--25]&0.0830(9)&-&0.1025(11)&-\\
 0.123000&[10--25]&-& 1.023(4)&-& 0.995(2)\\
 0.119457&[10--25]&-& 1.018(5)&-& 0.999(2)\\
 0.113638&[10--25]&-& 1.022(7)&-& 1.011(2)\\
 0.106031&[10--20]&-& 1.026(14)&-& 1.036(10)\\
 0.096555&[10--17]&-& 1.047(20)&-& 1.094(11)\\
\hline
\end{tabular}
\end{center}
\caption{\label{tabratE5}Heavy-strange meson masses and decay constants for the ensemble E5.}
\end{table}

\begin{table}[t]
\begin{center}
\begin{tabular}{|c|c|c|}
\hline
$\kappa_h$&$[t_{\rm min}$ -- $t_{\rm max}]$&$a m^{\rm PCAC}_{hs}$\\
\hline
  0.127130&[16--38]& 0.1354(8)\\
 0.123700&[16--38]& 0.1837(10)\\
 0.119241&[16--38]& 0.2495(13)\\
 0.113382&[16--25]& 0.3405(15)\\
 0.105793&[ 9--18]& 0.4824(10)\\
 0.096211&[ 9--16]& 0.6870(11)\\
 \hline
\end{tabular}
\end{center}
\caption{\label{tabmpcacF6} Average PCAC heavy and strange quark masses for the ensemble F6.}
\end{table}

\begin{table}[t]
\begin{center}
\begin{tabular}{|c|c|c|c|c|c|}
\hline
$\kappa_h$&$[t_{\rm min}$ -- $t_{\rm max}]$&$a M_{H_s}(i)$&$M_{H_s}(i+1)/M_{H_s}(i)$&$a M_{H^*_s}(i)$&$M_{H^*_s}(i+1)/M_{H^*_s}(i)$\\
\hline
0.127130&[10--42]& 0.6577(5)&-& 0.7058(9)&-\\
 0.123700&[10--42]& 0.7791(5)& 1.1847(2)& 0.8179(9)& 1.1589(5)\\
 0.119241&[10--28]& 0.9208(6)& 1.1818(3)& 0.9516(10)& 1.1635(4)\\
 0.113382&[10--26]& 1.0883(7)& 1.1819(2)& 1.1123(10)& 1.1688(3)\\
 0.105793&[10--23]& 1.2863(9)& 1.1819(2)& 1.3041(12)& 1.1725(4)\\
 0.096211&[10--17]& 1.5176(11)& 1.1798(3)& 1.5316(14)& 1.1744(6)\\
\hline
$\kappa_h$&$[t_{\rm min}$ -- $t_{\rm max}]$&$a f_{H_s}(i)$&$f_{H_s}(i+1)/f_{H_s}(i)$
&$a f_{H^*_s}(i)$&$f_{H^*_s}(i+1)/f_{H^*_s}(i)$\\
\hline
0.12713&[10--42]&0.0814(12)&-&0.0987(12)&-\\
 0.123700&[10--42]&-& 1.037(7)&-& 0.995(4)\\
 0.119241&[10--28]&-& 0.995(13)&-& 1.022(16)\\
 0.113382&[10--26]&-& 1.024(7)&-& 1.015(4)\\
 0.105793&[10--23]&-& 1.068(13)&-& 1.064(10)\\
 0.096211&[10--17]&-& 1.070(6)&-& 1.088(1)\\
\hline
\end{tabular}
\end{center}
\caption{\label{tabratF6}Heavy-strange meson masses and decay constants for the ensemble F6.}
\end{table}

\begin{table}[t]
\begin{center}
\begin{tabular}{|c|c|c|}
\hline
$\kappa_h$&$[t_{\rm min}$ -- $t_{\rm max}]$&$a m^{\rm PCAC}_{hs}$\\
\hline
0.127130&[16--40]& 0.1362(6)\\
 0.123649&[16--40]& 0.1847(8)\\
 0.119196&[16--36]& 0.2491(9)\\
 0.113350&[16--32]& 0.3391(11)\\
 0.105786&[ 9--27]& 0.4686(13)\\
 0.096689&[ 9--23]& 0.6505(17)\\
\hline
\end{tabular}
\end{center}
\caption{\label{tabmpcacF7}Average PCAC heavy and strange quark masses for the ensemble F7.}
\end{table}

\begin{table}[t]
\begin{center}
\begin{tabular}{|c|c|c|c|c|c|}
\hline
$\kappa_h$&$[t_{\rm min}$ -- $t_{\rm max}]$&$a M_{H_s}(i)$&$M_{H_s}(i+1)/M_{H_s}(i)$&$a M_{H^*_s}(i)$&$M_{H^*_s}(i+1)/M_{H^*_s}(i)$\\
\hline
0.127130&[10--41]& 0.6557(4)&-& 0.7032(10)&-\\
 0.123649&[10--41]& 0.7784(5)& 1.1872(2)& 0.8170(9)& 1.1617(3)\\
 0.119196&[10--40]& 0.9193(5)& 1.1810(2)& 0.9500(9)& 1.1628(2)\\
 0.113350&[10--35]& 1.0861(5)& 1.1814(1)& 1.1099(9)& 1.1683(2)\\
 0.105786&[10--28]& 1.2824(6)& 1.1808(1)& 1.3006(9)& 1.1718(2)\\
 0.096689&[10--26]& 1.5013(7)& 1.1707(1)& 1.5149(10)& 1.1648(2)\\
\hline
$\kappa_h$&$[t_{\rm min}$ -- $t_{\rm max}]$&$a f_{H_s}(i)$&$f_{H_s}(i+1)/f_{H_s}(i)$
&$a f_{H^*_s}(i)$&$f_{H^*_s}(i+1)/f_{H^*_s}(i)$\\
\hline
0.12713&[10--40]&0.0787(8)&-&0.0971(9)&-\\
0.123649&[10--40]&-& 1.010(4)&-& 0.994(2)\\
 0.119196&[10--40]&-& 1.003(6)&-& 0.997(2)\\
 0.113350&[10--35]&-& 1.019(8)&-& 1.001(7)\\
 0.105786&[10--28]&-& 1.047(16)&-& 1.041(8)\\
 0.096689&[10--26]&-& 1.056(16)&-& 1.083(7)\\
\hline
\end{tabular}
\end{center}
\caption{\label{tabratF7}Heavy-strange meson masses and decay constants for the ensemble F7.}
\end{table}

\begin{table}[t]
\begin{center}
\begin{tabular}{|c|c|c|}
\hline
$\kappa_h$&$[t_{\rm min}$ -- $t_{\rm max}]$&$a m_{\rm PCAC}(\kappa_h,\kappa_{\rm s})$\\
\hline
0.127100&[16--46]& 0.1374(5)\\
 0.123719&[16--46]& 0.1850(6)\\
 0.119260&[16--46]& 0.2502(7)\\
 0.113447&[16--36]& 0.3405(9)\\
 0.105836&[ 9--27]& 0.4710(11)\\
 0.096143&[ 9--23]& 0.6649(17)\\
\hline
\end{tabular}
\end{center}
\caption{\label{tabmpcacG8} Average PCAC heavy and strange quark masses for the ensemble G8.}
\end{table}

\begin{table}[t]
\begin{center}
\begin{tabular}{|c|c|c|c|c|c|}
\hline
$\kappa_h$&$[t_{\rm min}$ -- $t_{\rm max}]$&$a M_{H_s}(i)$&$M_{H_s}(i+1)/M_{H_s}(i)$&$a M_{H^*_s}(i)$&$M_{H^*_s}(i+1)/M_{H^*_s}(i)$\\
\hline
0.127100&[10--41]& 0.6568(4)&-& 0.7029(7)&-\\
 0.123719&[10--41]& 0.7762(5)& 1.1817(1)& 0.8140(7)& 1.1579(2)\\
 0.119260&[10--38]& 0.9174(4)& 1.1819(1)& 0.9477(6)& 1.1644(2)\\
 0.113447&[10--36]& 1.0833(6)& 1.1809(1)& 1.1072(6)& 1.1683(2)\\
 0.105836&[10--30]& 1.2808(6)& 1.1823(1)& 1.2993(6)& 1.1735(1)\\
 0.096143&[10--27]& 1.5141(7)& 1.1821(1)& 1.5278(7)& 1.1759(1)\\
\hline
$\kappa_h$&$[t_{\rm min}$ -- $t_{\rm max}]$&$a f_{H_s}(i)$&$f_{H_s}(i+1)/f_{H_s}(i)$
&$a f_{H^*_s}(i)$&$f_{H^*_s}(i+1)/f_{H^*_s}(i)$\\
\hline
0.12710&[11--41]&0.0800(7)&-&0.0954(6)&-\\
 0.123719&[11--41]&-& 1.015(3)&-& 0.995(1)\\
 0.119260&[11--38]&-& 1.006(5)&-& 1.000(3)\\
 0.113447&[11--36]&-& 1.005(8)&-& 1.007(3)\\
 0.105836&[11--30]&-& 1.015(11)&-& 1.043(6)\\
 0.096143&[11--21]&-& 1.038(24)&-& 1.085(6)\\
\hline
\end{tabular}
\end{center}
\caption{\label{tabratG8} Heavy-strange meson masses and decay constants for the ensemble G8.}
\end{table}

\begin{table}[t]
\begin{center}
\begin{tabular}{|c|c|c|}
\hline
$\kappa_h$&$[t_{\rm min}$ -- $t_{\rm max}]$&$a m^{\rm PCAC}_{hs}$\\
\hline
  0.130260&[16--42]& 0.0986(7)\\
 0.127737&[16--42]& 0.1336(8)\\
 0.124958&[16--42]& 0.1726(10)\\
 0.121051&[16--38]& 0.2288(12)\\
 0.115915&[16--36]& 0.3060(13)\\
 0.109399&[16--30]& 0.4117(15)\\
\hline
\end{tabular}
\end{center}
\caption{\label{tabmpcacN6} PCAC masses for the ensemble N6}
\end{table}

\begin{table}[t]
\begin{center}
\begin{tabular}{|c|c|c|c|c|c|}
\hline
$\kappa_h$&$[t_{\rm min}$ -- $t_{\rm max}]$&$a M_{H_s}(i)$&$M_{H_s}(i+1)/M_{H_s}(i)$&$a M_{H^*_s}(i)$&$M_{H^*_s}(i+1)/M_{H^*_s}(i)$\\
\hline
 0.130260&[10--42]& 0.4845(5)& -& 0.5216(6)&-\\
 0.127737&[10--42]& 0.5804(6)& 1.1978(2)& 0.6103(6)& 1.1700(3)\\
 0.124958&[10--42]& 0.6763(6)& 1.1652(2)& 0.7008(6)& 1.1484(2)\\
 0.121051&[10--40]& 0.7994(6)& 1.1821(2)& 0.8191(7)& 1.1688(2)\\
 0.115915&[10--36]& 0.9467(6)& 1.1842(2)& 0.9624(7)& 1.1749(2)\\
 0.109399&[10--34]& 1.1183(7)& 1.1812(2)& 1.1306(8)& 1.1747(1)\\
\hline
$\kappa_h$&$[t_{\rm min}$ -- $t_{\rm max}]$&$a f_{H_s}(i)$&$f_{H_s}(i+1)/f_{H_s}(i)$
&$a f_{H^*_s}(i)$&$f_{H^*_s}(i+1)/f_{H^*_s}(i)$\\
\hline
0.130260&[13--42]&0.0603(8)&-&0.0714(8)&-\\
  0.127737&[13--42]&-& 1.019(4)&-& 0.983(2)\\
 0.124958&[13--42]&-& 1.008(4)&-& 0.984(2)\\
 0.121051&[13--40]&-& 1.004(5)&-& 0.988(3)\\
 0.115915&[13--36]&-& 0.986(10)&-& 1.007(4)\\
 0.109399&[13--34]&-& 0.982(11)&-& 1.028(11)\\
 \hline
\end{tabular}
\end{center}
\caption{\label{tabratN6} Heavy-strange meson masses and decay constants for the ensemble N6.}
\end{table}

\begin{table}[t]
\begin{center}
\begin{tabular}{|c|c|c|}
\hline
$\kappa_h$&$[t_{\rm min}$ -- $t_{\rm max}]$&$a m_{\rm PCAC}(\kappa_h,\kappa_{\rm s})$\\
\hline
 0.130220&[16--46]& 0.0993(4)\\
 0.127900&[16--46]& 0.1315(5)\\
 0.124944&[16--46]& 0.1730(7)\\
 0.120910&[16--42]& 0.2309(9)\\
 0.115890&[16--40]& 0.3065(10)\\
 0.109400&[16--32]& 0.4118(14)\\
\hline
\end{tabular}
\end{center}
\caption{\label{tabmpcacO7} Average PCAC heavy and strange quark masses for the ensemble O7.}
\end{table}

\begin{table}[t]
\begin{center}
\begin{tabular}{|c|c|c|c|c|c|}
\hline
$\kappa_h$&$[t_{\rm min}$ -- $t_{\rm max}]$&$a M_{H_s}(i)$&$M_{H_s}(i+1)/M_{H_s}(i)$&$a M_{H^*_s}(i)$&$M_{H^*_s}(i+1)/M_{H^*_s}(i)$\\
\hline
0.130220&[10--55]& 0.4851(4)&-& 0.5213(6)&-\\
 0.127900&[10--55]& 0.5734(4)& 1.1821(2)& 0.6029(7)& 1.1566(3)\\
 0.124944&[10--55]& 0.6756(5)& 1.1782(2)& 0.6995(7)& 1.1602(2)\\
 0.120910&[10--50]& 0.8024(5)& 1.1877(2)& 0.8213(7)& 1.1740(2)\\
 0.115890&[10--46]& 0.9462(5)& 1.1792(2)& 0.9609(7)& 1.1701(2)\\
 0.109400&[10--34]& 1.1170(6)& 1.1805(2)& 1.1280(7)& 1.1739(2)\\
\hline
$\kappa_h$&$[t_{\rm min}$ -- $t_{\rm max}]$&$a f_{H_s}(i)$&$f_{H_s}(i+1)/f_{H_s}(i)$
&$a f_{H^*_s}(i)$&$f_{H^*_s}(i+1)/f_{H^*_s}(i)$\\
\hline
0.130220&[14--55]&0.0589(7)&-&0.0705(10)&-\\
 0.127900&[14--55]&-& 1.002(4)&-& 0.987(2)\\
 0.124944&[14--55]&-& 0.991(7)&-& 0.985(2)\\
 0.120910&[14--50]&-& 1.006(27)&-& 0.983(8)\\
 0.115890&[14--46]&-& 1.050(33)&-& 0.995(8)\\
 0.109400&[14--34]&-& 0.996(37)&-& 1.020(6)\\
 \hline
\end{tabular}
\end{center}
\caption{\label{tabratO7} Heavy-strange meson masses and decay constants for the ensemble O7.}
\end{table}


\newpage

\begin{thebibliography}{99}

\bibitem{LeesXJ}
  J.~P.~Lees {\it et al.} [BaBar Collaboration],
Phys.\ Rev.\ Lett.\  {\bf 109}, 101802 (2012).
[arXiv:1205.5442 [hep-ex]].

\bibitem{LeesUZD}
  J.~P.~Lees {\it et al.} [BaBar Collaboration],
Phys.\ Rev.\ D {\bf 88}, no. 7, 072012 (2013).
[arXiv:1303.0571 [hep-ex]].

\bibitem{HuschleRGA}
  M.~Huschle {\it et al.} [Belle Collaboration],
Phys.\ Rev.\ D {\bf 92}, no. 7, 072014 (2015).
[arXiv:1507.03233 [hep-ex]].

\bibitem{AaijYRA}
  R.~Aaij {\it et al.} [LHCb Collaboration],
Phys.\ Rev.\ Lett.\  {\bf 115}, no. 11, 111803 (2015), Erratum: [Phys.\ Rev.\ Lett.\  {\bf 115}, no. 15, 159901 (2015)].
[arXiv:1506.08614 [hep-ex]].

\bibitem{HiroseWFN}
  S.~Hirose {\it et al.} [Belle Collaboration],
Phys.\ Rev.\ Lett.\  {\bf 118}, no. 21, 211801 (2017).
[arXiv:1612.00529 [hep-ex]].

\bibitem{SatoSVK}
  Y.~Sato {\it et al.} [Belle Collaboration],
Phys.\ Rev.\ D {\bf 94}, no. 7, 072007 (2016).
[arXiv:1607.07923 [hep-ex]].

\bibitem{AbdesselamDGH}
  A.~Abdesselam {\it et al.} [Belle Collaboration],
[arXiv:1904.08794 [hep-ex]].

\bibitem{AaijORA}
  R.~Aaij {\it et al.} [LHCb Collaboration],
Phys.\ Rev.\ Lett.\  {\bf 113}, 151601 (2014).
[arXiv:1406.6482 [hep-ex]].

\bibitem{AaijVBB}
  R.~Aaij {\it et al.} [LHCb Collaboration],
JHEP {\bf 1708}, 055 (2017).
[arXiv:1705.05802 [hep-ex]].

\bibitem{AaijWAD}
  R.~Aaij {\it et al.} [LHCb Collaboration],
Phys.\ Rev.\ Lett.\  {\bf 122}, no. 19, 191801 (2019).
[arXiv:1903.09252 [hep-ex]].

\bibitem{AaijTYK}
  R.~Aaij {\it et al.} [LHCb Collaboration],
Phys.\ Rev.\ Lett.\  {\bf 120}, no. 12, 121801 (2018).
[arXiv:1711.05623 [hep-ex]].

\bibitem{BifaniZMI}
  S.~Bifani, S.~Descotes-Genon, A.~Romero Vidal and M.~H.~Schune,
J.\ Phys.\ G {\bf 46}, no. 2, 023001 (2019).
[arXiv:1809.06229 [hep-ex]].

\bibitem{DiLuzioVAT}
  L.~Di Luzio, A.~Greljo and M.~Nardecchia,
Phys.\ Rev.\ D {\bf 96}, no. 11, 115011 (2017).
[arXiv:1708.08450 [hep-ph]].

\bibitem{McNeileNG}
  C.~McNeile, C.~T.~H.~Davies, E.~Follana, K.~Hornbostel and G.~P.~Lepage,
Phys.\ Rev.\ D {\bf 85}, 031503 (2012).
[arXiv:1110.4510 [hep-lat]].

\bibitem{BazavovAA}
  A.~Bazavov {\it et al.} [Fermilab Lattice and MILC Collaborations],
Phys.\ Rev.\ D {\bf 85}, 114506 (2012).
[arXiv:1112.3051 [hep-lat]].

\bibitem{LinUR}
  H.~W.~Lin and N.~Christ,
Phys.\ Rev.\ D {\bf 76}, 074506 (2007).
[hep-lat/0608005].

\bibitem{ChristUS}
  N.~H.~Christ, M.~Li and H.~W.~Lin,
Phys.\ Rev.\ D {\bf 76}, 074505 (2007).
[hep-lat/0608006].

\bibitem{NaKP}
  H.~Na, C.~J.~Monahan, C.~T.~H.~Davies, R.~Horgan, G.~P.~Lepage and J.~Shigemitsu,
Phys.\ Rev.\ D {\bf 86}, 034506 (2012).
[arXiv:1202.4914 [hep-lat]].

\bibitem{DowdallTGA}
  R.~J.~Dowdall {\it et al.} [HPQCD Collaboration],
Phys.\ Rev.\ Lett.\  {\bf 110}, no. 22, 222003 (2013).
[arXiv:1302.2644 [hep-lat]].

\bibitem{BlossierMK}
  B.~Blossier {\it et al.} [ALPHA Collaboration],
JHEP {\bf 1012}, 039 (2010).
[arXiv:1006.5816 [hep-lat]].

\bibitem{DimopoulosGX}
  P.~Dimopoulos {\it et al.} [ETM Collaboration],
JHEP {\bf 1201}, 046 (2012).
[arXiv:1107.1441 [hep-lat]].

\bibitem{CarrascoZTA}
  N.~Carrasco {\it et al.} [ETM Collaboration],
JHEP {\bf 1403}, 016 (2014).
[arXiv:1308.1851 [hep-lat]].

\bibitem{BernardoniFVA}
  F.~Bernardoni {\it et al.} [ALPHA Collaboration],
Phys.\ Lett.\ B {\bf 735}, 349 (2014).
[arXiv:1404.3590 [hep-lat]].


\bibitem{AokiNGA}
  Y.~Aoki, T.~Ishikawa, T.~Izubuchi, C.~Lehner and A.~Soni,
Phys.\ Rev.\ D {\bf 91}, no. 11, 114505 (2015).
[arXiv:1406.6192 [hep-lat]].

\bibitem{ChristUEA}
  N.~H.~Christ, J.~M.~Flynn, T.~Izubuchi, T.~Kawanai, C.~Lehner, A.~Soni, R.~S.~Van de Water and O.~Witzel,
Phys.\ Rev.\ D {\bf 91}, no. 5, 054502 (2015).
[arXiv:1404.4670 [hep-lat]].

\bibitem{BussoneIUA}
  A.~Bussone {\it et al.} [ETM Collaboration],
Phys.\ Rev.\ D {\bf 93}, no. 11, 114505 (2016).
[arXiv:1603.04306 [hep-lat]].

\bibitem{BazavovLYH}
  A.~Bazavov {\it et al.},
Phys.\ Rev.\ D {\bf 98}, no. 7, 074512 (2018).
[arXiv:1712.09262 [hep-lat]].

\bibitem{HughesSPC}
  C.~Hughes, C.~T.~H.~Davies and C.~J.~Monahan,
Phys.\ Rev.\ D {\bf 97}, no. 5, 054509 (2018).
[arXiv:1711.09981 [hep-lat]].


\bibitem{BoyleKNM}
  P.~A.~Boyle {\it et al.} [RBC/UKQCD Collaboration],
[arXiv:1812.08791 [hep-lat]].

\bibitem{ColquhounOHA}
  B.~Colquhoun {\it et al.} [HPQCD Collaboration],
Phys.\ Rev.\ D {\bf 91}, no. 11, 114509 (2015).
[arXiv:1503.05762 [hep-lat]].

\bibitem{LubiczASP}
  V.~Lubicz {\it et al.} [ETM Collaboration],
[arXiv:1707.04529 [hep-lat]].

\bibitem{MelikhovYU}
  D.~Melikhov and B.~Stech,
Phys.\ Rev.\ D {\bf 62}, 014006 (2000).
[hep-ph/0001113].

\bibitem{EbertHJ}
  D.~Ebert, R.~N.~Faustov and V.~O.~Galkin,
Phys.\ Lett.\ B {\bf 635}, 93 (2006).
[hep-ph/0602110].

\bibitem{GelhausenWIA}
  P.~Gelhausen, A.~Khodjamirian, A.~A.~Pivovarov and D.~Rosenthal,
Phys.\ Rev.\ D {\bf 88}, 014015 (2013), Erratum: [Phys.\ Rev.\ D {\bf 89}, 099901 (2014)], Erratum: [Phys.\ Rev.\ D {\bf 91}, 099901 (2015)].
[arXiv:1305.5432 [hep-ph]].

\bibitem{NarisonSKA}
  S.~Narison,
Int.\ J.\ Mod.\ Phys.\ A {\bf 30}, no. 20, 1550116 (2015).
[arXiv:1404.6642 [hep-ph]].

\bibitem{LuchaXUA}
  W.~Lucha, D.~Melikhov and S.~Simula,
Phys.\ Rev.\ D {\bf 91}, no. 11, 116009 (2015).
[arXiv:1504.03017 [hep-ph]].

\bibitem{WangMXA}
  Z.~G.~Wang,
Eur.\ Phys.\ J.\ C {\bf 75}, 427 (2015).
[arXiv:1506.01993 [hep-ph]].

\bibitem{BroadhurstSE}
  D.~J.~Broadhurst and A.~G.~Grozin,
Phys.\ Rev.\ D {\bf 52}, 4082 (1995).
[hep-ph/9410240].

\bibitem{ChetyrkinVI}
  K.~G.~Chetyrkin and A.~G.~Grozin,
Nucl.\ Phys.\ B {\bf 666}, 289 (2003).
[hep-ph/0303113].

\bibitem{ChetyrkinPQ}
  K.~G.~Chetyrkin and A.~Retey,
Nucl.\ Phys.\ B {\bf 583}, 3 (2000).
[hep-ph/9910332].

\bibitem{BekavacZC}
  S.~Bekavac, A.~G.~Grozin, P.~Marquard, J.~H.~Piclum, D.~Seidel and M.~Steinhauser,
Nucl.\ Phys.\ B {\bf 833}, 46 (2010).
[arXiv:0911.3356 [hep-ph]].


\bibitem{BlossierHG}
  B.~Blossier {\it et al.} [ETM Collaboration],
JHEP {\bf 1004}, 049 (2010).
[arXiv:0909.3187 [hep-lat]].


\bibitem{BernardoniXBA}
  F.~Bernardoni {\it et al.},
Phys.\ Lett.\ B {\bf 730}, 171 (2014).
[arXiv:1311.5498 [hep-lat]].

\bibitem{FritzschAW}
  P.~Fritzsch, J.~Heitger and N.~Tantalo,
JHEP {\bf 1008}, 074 (2010).
[arXiv:1004.3978 [hep-lat]].

\bibitem{SheikholeslamiIJ}
  B.~Sheikholeslami and R.~Wohlert,
Nucl.\ Phys.\ B {\bf 259}, 572 (1985).

\bibitem{LuscherUG}
  M.~L\"uscher, S.~Sint, R.~Sommer, P.~Weisz and U.~Wolff,
Nucl.\ Phys.\ B {\bf 491}, 323 (1997).
[hep-lat/9609035].

\bibitem{WilsonSK}
  K.~G.~Wilson,
Phys.\ Rev.\ D {\bf 10}, 2445 (1974).

\bibitem{LottiniRFA}
  S.~Lottini [ALPHA Collaboration],
PoS LATTICE {\bf 2013}, 315 (2014).
[arXiv:1311.3081 [hep-lat]].


\bibitem{BlossierJOL}
  B.~Blossier, J.~Heitger and M.~Post,
Phys.\ Rev.\ D {\bf 98}, no. 5, 054506 (2018).
[arXiv:1803.03065 [hep-lat]].

\bibitem{DellaMorteDYU}
  M.~Della Morte {\it et al.},
JHEP {\bf 1710}, 020 (2017).
[arXiv:1705.01775 [hep-lat]].

\bibitem{DellaMorteAQE}
  M.~Della Morte, R.~Hoffmann and R.~Sommer,
JHEP {\bf 0503}, 029 (2005).
[hep-lat/0503003].

\bibitem{AokiCCA}
  S.~Aoki {\it et al.} [Flavour Lattice Averaging Group],
[arXiv:1902.08191 [hep-lat]].

\bibitem{BernardoniNQA}
  F.~Bernardoni {\it et al.},
Phys.\ Rev.\ D {\bf 92}, no. 5, 054509 (2015).
[arXiv:1505.03360 [hep-lat]].

\bibitem{BecirevicKAA}
  D.~Becirevic, A.~Le Yaouanc, A.~Oyanguren, P.~Roudeau and F.~Sanfilippo,
[arXiv:1407.1019 [hep-ph]].
 
\end{thebibliography}
\end{document}